\RequirePackage{fix-cm}
\documentclass{iopjournal}
\usepackage{amsmath,amssymb,amsfonts}
\usepackage[sort]{cite}

\newcommand{\de}{\text{d}}

\newcommand{\AdSST}{\text{AdS}_3\times\text{S}^3\times\text{T}^4}
\newcommand{\Xd}{\dot{X}}
\newcommand{\Xp}{\acute{X}}
\newcommand{\xp}{\acute{x}}

\begin{document}

\articletype{Paper} 

\title{Tree-level S matrix for \texorpdfstring{$\lambda$}{lambda}-deformed AdS\texorpdfstring{$_3$}{3} strings}

\author{Marco Costantino$^{a}$\orcid{0009-0000-6668-2102}, 
Silvia Penati$^{a,b}$\orcid{0000-0003-0733-8146}}, Alessandro Sfondrini$^{c,d}$\footnote{Corresponding author. On leave from the University of Padova, Italy.}\orcid{0000-0001-5930-3100} 

\affil{$^a$Dipartimento di Fisica, Universit\`a degli Studi di Milano–Bicocca, Piazza della Scienza 3, 20126 Milano, Italy}

\affil{$^b$INFN, Sezione di Milano–Bicocca, Piazza della Scienza 3, 20126 Milano, Italy}

\affil{$^c$School of Mathematics, University of Birmingham,
Watson Building, Edgbaston, Birmingham B15 2TT, UK}

\affil{$^d$INFN, Sezione di Padova, via Marzolo 8, 35131 Padova, Italy}

\email{m.costantino7@campus.unimib.it, silvia.penati@mib.infn.it, a.sfondrini@bham.ac.uk}

\begin{abstract}
\begin{minipage}{\textwidth}
We consider the supersymmetric $\lambda$-deformation of $\text{AdS}_3 \times \text{S}^3 \times \text{T}^4$ superstrings and compute its perturbative bosonic tree-level worldsheet S matrix in the light-cone gauge. For generic values of $0 \leq \lambda < 1$, we show that the worldsheet scattering remains compatible with integrability due to a non-trivial cancellation of non-elastic scattering processes. By contrast, the S matrix becomes ill-defined for $\lambda \to 1$, despite the fact that this limit reproduces the non-Abelian T-dual geometry up to an analytic continuation. This suggests that 
the $\lambda \to 1$ limit does not capture the full worldsheet dynamics of the T-dual theory.
\end{minipage}
\end{abstract}
\vspace{\baselineskip}

\vspace{10pt}
\noindent\rule{\textwidth}{1pt}
\tableofcontents
\noindent\rule{\textwidth}{1pt}
\vspace{10pt}

\section{Introduction}
\label{sec:intro}

\vskip 6pt
A notorious difficulty in understanding string theory on curved backgrounds is that, in most cases, it is not known how to quantise the string action. This is due to the presence of Ramond-Ramond (RR) background fluxes, which substantially complicate the worldsheet theory, making the Ramond--Neveu--Schwarz action non-local. Unfortunately, RR fluxes appear in almost all curved-space superstring backgrounds, including those of interest for the AdS/CFT correspondence~\cite{Maldacena:1997re}.
There are mainly two exceptions: certain $\text{AdS}_3$ backgrounds that can be supported by Neveu-Schwarz-Neveu-Schwarz (NSNS) fluxes only, and can be quantised as Wess-Zumino-Witten (WZW) models based on the $SL(2,\mathbb{R})$ algebra~\cite{Maldacena:2000hw}, and pp-wave backgrounds~\cite{Blau:2002dy}. For the latter, the RR flux is constant and the superstring spectrum can be worked out in the light-cone gauge, where the model becomes free~\cite{Metsaev:2002re}.

In general, away from these very special setups, how to quantise the superstring (or at least its spectrum) is still an open question.
However, for a surprisingly large family of superstring backgrounds with RR flux a route to quantisation and to efficiently computing the spectrum is provided by \textit{integrability}. This is the observation that, if the Green--Schwarz action defines a classically integrable model, one may try to exploit the symmetries of integrability to quantise the model in the light-cone gauge. In a sense, this approach builds on that of pp-wave strings, with the crucial difference that, after gauge-fixing, the worldsheet theory is far from being free, and quantising it without integrability would be hopeless. This approach was pioneered for $\text{AdS}_5\times\text{S}^5$ type-IIB superstrings, see~\cite{Arutyunov:2009ga,Beisert:2010jr} for reviews.

The integrability approach to the spectral problem is somewhat more convoluted than the worldsheet-CFT one.
This is likely due to the fact that the spectrum of strings in RR backgrounds is itself much more involved. For instance, the spectrum of the $SL(2,\mathbb{R})$ stringy WZW model~\cite{Maldacena:2000hw} can be written in closed form and is very degenerate, while the spectrum of $\text{AdS}_5\times\text{S}^5$ superstrings is non-degenerate and cannot be written in terms of elementary or special functions as a function of the string tension. Indeed, the latter spectrum is as complicated as the spectrum of a planar four-dimensional gauge theory.
The integrability approach then requires simplifying the problem by first studying the model (here, the light-cone gauge worldsheet theory) in a decompactification limit, where the crucial object is the worldsheet S~matrix~\cite{Zamolodchikov:1978xm}, and then finding the spectrum by thermodynamic Bethe ansatz (TBA) techniques~\cite{Zamolodchikov:1989cf} --- or more precisely, ``mirror'' TBA in the case of the superstring~\cite{Arutyunov:2007tc}.

One might expect that integrable superstring backgrounds would be few and far between, and very special (e.g., very supersymmetric). Instead, it turns out that there exist \textit{multi-parametric families} of integrable backgrounds, with little or no supersymmetry. A powerful way to construct such families relies on \textit{integrable deformations} of the simplest integrable backgrounds, see  e.g.~\cite{Hoare:2021dix} for a review. In this way, it is possible to break some or all of the (super)symmetries of the background, while preserving integrability.

In this paper we are interested in a particular deformation of a certain $\text{AdS}_3$ background. As already mentioned, $\text{AdS}_3$ backgrounds are special, and their deformations are particularly rich, see~\cite{Seibold:2024qkh} for a recent review. We will consider the simplest $\text{AdS}_3$ background, which is $\text{AdS}_3\times\text{S}^3\times\text{T}^4$.
This background preserves 16 Killing spinors (half of the maximal amount), and it can be supported by \textit{an arbitrary mixture} of NSNS and RR fluxes (more, specifically, three-form fluxes), reducing to a WZW model at the special pure-NSNS point, see~\cite{Demulder:2023bux} for a detailed review. Moreover, its symmetries take a factorised form,  $\mathfrak{psu}(1,1|2)\oplus\mathfrak{psu}(1,1|2)$,%
\footnote{The bosonic part of the $\mathfrak{psu}(1,1|2)$ Lie superalgebra is given by a non-compact $\mathfrak{su}(1,1)\cong\mathfrak{sl}(2,\mathbb{R})$ and a compact $\mathfrak{su}(2)$ R-symmetry. Indeed, the bosonic isometries of the background are $\mathfrak{so}(2,2)\cong\mathfrak{sl}(2,\mathbb{R})\oplus\mathfrak{sl}(2,\mathbb{R})$ and $\mathfrak{so}(4)\cong\mathfrak{su}(2)\oplus\mathfrak{su}(2)$.}
which allows one to deform either copy separately, as was done in~\cite{Hoare:2014oua,Sfetsos:2017sep,Seibold:2019dvf,Hoare:2022asa,Hoare:2022vnw}. The upshot of considering such a deformation is that the resulting family of backgrounds manifestly preserves 8 Killing spinors, and it turns out to have a better-behaved (less singular) metric with respect to the other similar deformations. More complicated backgrounds which have the same amount of supersymmetry are $\text{AdS}_3\times\text{S}^3\times\text{K3}$ and $\text{AdS}_3\times\text{S}^3\times\text{S}^3\times\text{S}^1$. The latter admits a particularly rich family of deformations, see e.g.~\cite{Seibold:2025fnu} and references therein, as well as the recent works~\cite{Maurelli:2025iba,Maurelli:2025ueo,Maurelli:2026dpp}.

The type of deformation that we consider here belongs to the class of ``$\lambda$ deformations''. Such deformations were initially introduced for the principal chiral model (PCM)~\cite{Sfetsos:2013wia}, then generalized to symmetric~\cite{Hollowood:2014rla,Sfetsos:2014cea} and semi-symmetric sigma models~\cite{Hollowood:2014qma}. 
In general, they are believed to deform the underlying integrability structure to that of a quantum group, whose quantum parameter $q$ has $|q|=1$.%
\footnote{Another class of quantum deformation goes under the name of $\eta$ deformations~\cite{Arutyunov:2013ega}, which emerge from ``Yang--Baxter'' deformations~\cite{Klimcik:2002zj,Delduc:2013qra,Kawaguchi:2014qwa,vanTongeren:2015soa}, and are believed to be related to the case $q\in\mathbb{R}$~\cite{Delduc:2017brb}. $\lambda$ and $\eta$ deformations may be related to each other by a Poisson-Lie T-duality supplemented by analytic continuation~\cite{Sfetsos:2015nya,Hoare:2014pna,Hoare:2015gda,Vicedo:2015pna}.}
Owing to the factorised form of the $\mathfrak{psu}(1,1|2)\oplus\mathfrak{psu}(1,1|2)$ symmetry, one can consider a ``bi-$\lambda$ deformation'' where each copy gets its own deformation parameter $\lambda_1,\lambda_2$~\cite{Sfetsos:2014cea,Sfetsos:2017sep,Hoare:2022asa}. To preserve half of the supersymmetry one sets e.g.~$\lambda_2=0$, and is left with a single parameter~$\lambda_1\equiv\lambda$, as it was done in~\cite{Hoare:2022asa,Hoare:2022vnw}. In this paper we will indeed focus on the one-parameter case, while leaving one of the two $\mathfrak{psu}(1,1|2)$ algebras untouched.
This setup was also considered in~\cite{Itsios:2023kma}, where is was shown that for any value of $0\leq\lambda<1$, one finds a consistent superstring background (i.e., NSNS and RR fluxes may be written down in such a way to solve the supergravity equations).
At $\lambda=0$, the starting point of the deformation, we are dealing with the pure-NSNS $\text{AdS}_3\times\text{S}^3\times\text{T}^4$ background --- the one which corresponds to a WZW model in the conformal gauge. When $0<\lambda<1$, it is necessary to introduce RR fluxes, and the geometry becomes more complicated (part of the isometries of $\text{AdS}_3$ and of the three-sphere are broken). The limit $\lambda\to1$ is singular but, as explained in~\cite{Itsios:2023kma}, can be related to the non-Abelian T-dual of $\text{AdS}_3\times\text{S}^3\times\text{T}^4$ by rescaling the coordinates and performing an analytic continuation (see also~\cite{Sfetsos:2013wia}). Hence, this particular deformation produces a supersymmetric superstring background that interpolates between particularly interesting models, and this is what makes it especially interesting for us.

A natural question is what is the worldsheet S~matrix for the light-cone gauge-fixed model when $\lambda>0$. Ideally, we would like to determine it exactly as a function of the various parameters of the theory, but this can be difficult for such deformed backgrounds.%
\footnote{%
The S matrix of the undeformed $\AdSST$ background has been worked out in~\cite{Borsato:2014hja,Lloyd:2014bsa,Dei:2018mfl}, and that of $\eta$-deformed $\AdSST$ superstrings was studied in \cite{Hoare:2014oua}. 
The study of the S~matrix of $\lambda$-deformed strings is still ongoing~\cite{Appadu:2017bnv}.
}
An important intermediate step is to determine the \textit{perturbative} S~matrix, obtained by expanding around the pp-wave case~\cite{Berenstein:2002jq}, order by order in the (large) string tension. Consistency with such a perturbative computation it is a crucial test for any exact S~matrix.%
\footnote{For the $\eta$-deformed $\AdSST$ superstring the perturbative computation was carried out in~\cite{Seibold:2021lju}.}
For the case at hand, the supersymmetric $\lambda$-deformation, a perturbative computation constitutes a check of integrability (which is expected to hold at the classical level, at least before gauge-fixing~\cite{Sfetsos:2017sep}). If the model is integrable, we expect the scattering to be \textit{factorised} and \textit{purely elastic}. For the two-to-two perturbative S~matrix $S(p_1,p_2)$ this leads to two requirements: first, $S(p_1,p_2)$ should satisfy the Yang--Baxter equation (or its perturbative expansion, order by order in the string tension); second, scattering two particles of momentum $p_1$ and $p_2$ must result in two particles of momentum $p_2$ and $p_1$ --- the very same momenta, just in the opposite order. The latter requirement is trivially enforced by kinematics for a simple relativistic model, but this is not the case on the string worldsheet --- in this case, \textit{not even at tree level} as it can be seen from the pp-wave analysis of~\cite{Itsios:2023kma}.
Hence, computing the tree-level S~matrix for supersymmetric $\lambda$-deformed $\AdSST$ superstring, provides both a test of (classical) integrability and a benchmark for future  studies of this model. This is the goal of this work.

The paper is structured as follows. In section~\ref{sec:lambda-def} we summarise the model following~\cite{Hoare:2022asa,Itsios:2023kma}. In section~\ref{sec:lcgauge} we determine the Hamiltonian in the light-cone gauge, up to quartic order in the bosonic transverse modes of the string. In section~\ref{sec:Hamiltonian} we work out the free-particle spectrum.  Sections~\ref{sec:Smatrix}~and~\ref{sec:Smatrixresults} are the core of the paper, where we compute the tree-level bosonic S~matrix and summarize our results, discussing in particular their compatibility with integrability. In section \ref{sec:SmatrixNATD} we discuss the S matrix in the $\lambda\to 1$ limit and we conclude in section~\ref{sec:conclusions}.

\section{The \texorpdfstring{$\lambda$}{lambda}-deformed model}
\label{sec:lambda-def}

\vskip 6pt
The computation of the worldsheet S~matrix for $\AdSST$ (super)strings was reviewed in~\cite{Demulder:2023bux}. Following the conventions introduced there, we consider the bosonic string action
\begin{equation}
\label{eq:stringaction}
    S_{\text{bos}}=-\frac{T}{2}\int\limits_{-\infty}^{+\infty}\de\tau \int\limits_{0}^{R}\de\sigma\left(\sqrt{-h}h^{ab}G_{\mu\nu}(X)+\varepsilon^{ab}B_{\mu\nu}(X)\right) \partial_a X^\mu\partial_b X^\nu
\end{equation}
on a cylinder of unit radius, where the ten coordinates of the (deformed) $\AdSST$ geometry are indicated as $X^\mu$.
Here $T$ denotes the string tension, expressed in dimensionless units.
The $\lambda$-deformed metric is given by~\cite{Itsios:2023kma} 
\begin{equation}
\label{eq:Metric}
\begin{aligned}
 \de s^2=G_{\mu\nu}(X)\de X^\mu\de X^\nu=&\frac{1+\lambda}{1-\lambda}\de\tilde{\alpha}^2+\ \frac{1-\lambda^2}{\tilde{\Delta}}\cosh^2\tilde{\alpha}(\de\tilde{\beta}^2-\cosh^2\tilde{\beta}\ \de\tilde{\gamma}^2)+ \\ &+
    \frac{1+\lambda}{1-\lambda}\de\alpha^2+\frac{1-\lambda^2}{\Delta}\text{sin}^2\alpha(\de\beta^2+\text{sin}^2\beta\ \de\gamma^2)+\sum_{i=5}^8 \de x_i^2 ,
\end{aligned}
\end{equation} 
while the Kalb--Ramond two-form is
\begin{equation}
     B=\biggr[\tilde{\alpha}+\frac{(1-\lambda)^2}{\tilde{\Delta}}\cosh\tilde{\alpha}\ \text{sinh}\tilde{\alpha}\biggr]\cosh\tilde{\beta} \, \de\tilde{\beta} \wedge \de\tilde{\gamma}+\biggr[-\alpha+\frac{(1-\lambda)^2}{\Delta}\cos\alpha\ \text{sin}\alpha\biggr]\text{sin}\beta \, \de\beta \wedge \de\gamma\,.
\end{equation}
These quantities are written using the shorthands
\begin{equation}
\tilde{\Delta}=(1+\lambda)^2\cosh^2\tilde{\alpha}-(1-\lambda)^2\text{sinh}^2\tilde{\alpha}\,,\qquad
    \Delta = (1-\lambda)^2\cos^2\alpha+(1+\lambda)^2\text{sin}^2\alpha\,.
\end{equation}
For this background to solve the supergravity equations, it is necessary to introduce a dilaton
\begin{equation}
    e^{-2\Phi}=\frac{\Delta\tilde{\Delta}}{(1+\lambda)^4}\,,
\end{equation}
as well as $F_1$, $F_3$ and $F_5$ RR fluxes, whose explicit expressions can be found in~\cite{Itsios:2023kma}.
In what follows we will be considering the tree-level worldsheet S~matrix for the bosonic excitations. Therefore, we can neglect the contribution of the Fradkin--Tseytlin term due to non-constant dilaton, and we will not need the explicit form of the RR fluxes.

\paragraph{The underformed model.}
Setting $\lambda=0$, we recover the undeformed $\AdSST$ metric (with unit radius), along with a non-trivial $B$-field. At $\lambda=0$ one also finds $F_1=F_3=F_5=0$~\cite{Itsios:2023kma}, and the dilaton is constant. Indeed, in this case the action~\eqref{eq:stringaction} in the conformal gauge becomes that of a WZW model. Since the supergravity equations force the radius of AdS$_3$ to be equal to the one of the three-sphere, in our conventions we find that the tension is quantised as $T=k/2\pi$, with $k$ integer.

\paragraph{The non-abelian T-dual.}
Turning on $\lambda>0$, we see that the metric degenerates as $\lambda\to1$. As discussed in~\cite{Itsios:2023kma} (see also~\cite{Sfetsos:2013wia}), through an analytic continuation and a suitable rescaling of the coordinates
\begin{equation}
    \begin{aligned}
        \tilde \alpha=i\frac{\pi}{2}+\frac{\rho}{2T}, \qquad \alpha=\frac{r}{2T},  \qquad \lambda=1-\frac{1}{T},
    \end{aligned}
    \label{eq:NATD_changeofcoordinates}
\end{equation}
supported by the $T\to \infty$ limit, it is possible to relate the $\lambda\to1$ limit of this metric to the one of the non-abelian T-dual of the principal chiral model on $\AdSST$. In fact, with suitable rescalings of the fluxes it is also possible to obtain the full supergravity solution for the non-abelian T-dual, as detailed in~\cite{Itsios:2023kma}.

\section{The light-cone Hamiltonian}
\label{sec:lcgauge}

\vskip 6pt
We are interested in quantising the action~\eqref{eq:stringaction} in a suitable uniform light-cone gauge~\cite{Arutyunov:2005hd} to extract the tree-level S~matrix on the worldsheet, see~\cite{Arutyunov:2009ga,Demulder:2023bux} for a review. We will be mainly interested in the cases with $0<\lambda<1$, i.e.\ generically deformed backgrounds.

\subsection{\bf Near-pp-wave geometry}
\label{subsection:near-pp-wave geometry}

\vskip 6pt
Following~\cite{Itsios:2023kma} we briefly discuss the pp-limit of the metric \eqref{eq:Metric} along a null geodesics.

Requiring no acceleration in both the AdS$_3$ and the S$^3$ directions ($\partial_\tau f(\tau)=\dot f(\tau)$),
\begin{equation}
    \ddot\alpha(\tau)=\ddot\beta(\tau)=\ddot\gamma(\tau)=0, \,\qquad \ddot{\tilde\alpha}(\tau)=\ddot{\tilde\beta}(\tau)=\ddot{\tilde\gamma}(\tau)=0,
\end{equation}
the geodesics equations read
\begin{equation}
    \begin{aligned}
        \left(\dot{\tilde\beta}-\cosh^2\tilde\beta\,\dot{\tilde\gamma}\right)\sinh(2\tilde\alpha)&=0,
        \, \qquad 
         &\left(\dot{\beta}+\sin^2\beta\,\dot{\gamma}\right)\sin(2\alpha)=0,
        \, 
        \\
        \sinh(2\tilde\beta)\dot{\tilde{\gamma}}^2+4\dot{\tilde\alpha}\dot{\tilde\beta}\,\frac{(\lambda-1)^2}{\tilde\Delta}\tanh{\tilde\alpha\,}&=0,
        \,\qquad
        &\sin(2\beta)\dot{{\gamma}}^2-4\dot{\alpha}\dot{\beta}\,\frac{(\lambda-1)^2}{\Delta}\cot{\alpha\,}=0,
        \\
        \left(\dot{\tilde\beta}\,\tanh\tilde{\beta}+\frac{(\lambda-1)^2}{\tilde\Delta}\dot{\tilde\alpha}\tanh(\tilde\alpha)\right)\dot{\tilde\gamma}&=0,
        \, \qquad
        &\left(\dot{\beta}\,\cot{\beta}+\frac{(\lambda-1)^2}{\Delta}\dot{\alpha}\cot(\alpha)\right)\dot{\gamma}&=0.
    \end{aligned}
\end{equation}
It is natural to consider a null geodesic along the two isometric coordinates $\gamma$ and $\tilde{\gamma}$ with non-vanishing velocity. Due to the highly constrained form of the equations, the only allowed solutions are
\begin{gather}
    \begin{cases}
        \tilde\alpha(\tau)=\tilde\beta(\tau)=0
        \\
       \alpha(\tau)=\beta(\tau)=\frac{\pi}{2}
    \end{cases}
    \qquad \text{and} \qquad
    \begin{cases}
       \tilde\alpha(\tau)=\tilde\beta(\tau)=0
        \\
         \alpha(\tau)=\beta(\tau)=0
    \end{cases}.
\end{gather}
However, the solution $\beta(\tau)=0$ is not acceptable, as it shrinks the $\gamma$ direction to zero. We are then left with 
\begin{equation}
    \tilde{\gamma}(\tau)=\gamma(\tau)=c\,\tau\,,\qquad
    \tilde{\alpha}(\tau)=\tilde{\beta}(\tau)=0\,,\qquad
    \alpha(\tau)=\beta(\tau)=\frac{\pi}{2}\, ,
    \label{eq:geodesic}
\end{equation}
where $c$ is the velocity along the two directions. 

In order to write the metric in an adapted set of coordinates, 
it is convenient to define the light-cone coordinates $X^\pm$ as
\begin{equation}
\tilde{\gamma} = X^+-a\,X^-, \qquad 
\gamma = X^+ + (\kappa^2-a)\, X^-\,,
\label{eq:ppwave_changeofcoord_LC}
\end{equation}
such that on the chosen null geodesics we have
\begin{equation}
    X^-=\frac{\gamma-\tilde\gamma}{\kappa^2}=0, \qquad \text{and} \ \ X^+\cdot X^-=0,
\end{equation}
and the four transverse ones, $x_i, i=1,2,3,4$ as 
\begin{equation}
\tilde{\alpha} = \frac{1}{\kappa} x_1\,, \qquad 
\alpha = \frac{\pi}{2} + \frac{1}{\kappa} \, x_2\,,
\qquad
\tilde{\beta} = \kappa \, x_3\,, \qquad 
\beta = \frac{\pi}{2} + \kappa \, x_4\,,
\label{eq:ppwave_changeofcoord}
\end{equation}
where we have defined $\kappa=\sqrt{\frac{1+\lambda}{1-\lambda}}$,
and $a$ is a generic real parameter which parametrizes the most general gauge such that the conjugate momentum $P_+$ is equal to $P_{\tilde\gamma}+P_\gamma$. Different choices of $a$ correspond to different definitions of the coordinate and can possibly simplify the computations. 

In terms of these new coordinates the line element takes the following form
\begin{equation}
\begin{aligned}
    \de s^2&=\de x_1^2+\frac{1-\lambda^2}{\tilde{\Delta}}\cosh^2(x_1/\kappa)\left[\kappa\, \de x_3^2-\cosh^2(\kappa\, x_3)(\de X^+-a\, \de X^-)^2\right]
    \\
    &\quad+\de x_2^2+\frac{1-\lambda^2}{\Delta}\cos^2(x_2/\kappa)\left[\kappa\, \de x_4^2+\cos^2(\kappa x_4)  (\de X^++(\kappa^2-a)\,\de X^-)^2\right]+\sum_{j=5}^8\de x_i^2\, .
    \label{eq:deformed_metric}
\end{aligned}
\end{equation}

Moreover, it is convenient to modify the Kalb--Ramond field by adding a closed term, so that we have
\begin{equation}
\label{eq:Bfield}
\begin{aligned}
    B&=\frac{1}{2}\cosh(\kappa\,x_3)\Bigl(\kappa\,x_1+\frac{1-\lambda^2}{2\tilde \Delta}\sinh{(2\,x_1/\kappa})\Bigl)\,\de x_3\wedge (\de X^+-a\,\de X^-)\\ 
    &\quad
    -\frac{1}{2}\sinh{(\kappa\,x_3)}\Bigl(\frac{1}{\kappa}+\frac{(1-\lambda)^2}{\tilde\Delta^2}((1+\lambda^2)\cosh{(2\,x_1/\kappa)+2\lambda})\Bigl)\, \de x_1\wedge (\de X^+-a\,\de X^-)
    \\
    &\quad
    +\frac{1}{2}\kappa\cos(\kappa\,x_4)\Bigl(-\kappa\,x_2-\frac{1-\lambda^2}{2\Delta}\sin(2\,x_2/\kappa)\Bigl)
\, \de x_4\wedge(\de X^++(\kappa^2-a)\,\de X^-) 
    \\
    &\quad
    -\frac{1}{2}\sin(\kappa\  x_4)\Bigl(-\frac{1}{\kappa}-\frac{(1-\lambda)^2}{\Delta^2}((1+\lambda^2)\cos(2\,x_2/\kappa)+2\lambda)\Bigl)\, \de x_2\wedge (\de X^++(\kappa^2-a)\,\de X^-) \, .
\end{aligned}
\end{equation}

With the particular choice of the coefficients in definitions \eqref{eq:ppwave_changeofcoord_LC} and \eqref{eq:ppwave_changeofcoord}, and an appropriate rescaling\footnote{Precisely, we rescale $X^- \to \frac{1}{T} X^-$, $x^i \to \frac{1}{\sqrt{T}} x^i$, and include the $T$ factor in front of the action.} of the coordinates by powers of the string tension $T$, we obtain an expansion of the metric in powers of $1/T$ around the pp-wave geometry. Taking the $T \to \infty$ limit, the pp-wave line element reads
\begin{equation}
    \de s^2=2\de X^+\de X^-+A(x_i)\,\de X^{+2}+\sum_{i=1}^8\de x_i^2 \, .
    \label{eq:metric_ppwave_generic}
\end{equation}

The string spectrum on the pp-wave background can be solved exactly in the light-cone gauge, because the resulting action is quadratic in the fields. We are however interested in the full $\lambda$-deformed background, not just its pp-wave limit. In what follows, we will keep the subleading terms in the $1/T$ expansion around the pp-wave background. Then, the quartic terms in the gauge-fixed action will give rise to the perturbative worldsheet S~matrix of the $\lambda$-deformed backgrounds.

\subsection{\bf The Hamiltonian expansion}

\vskip 6pt
We consider the string action~\eqref{eq:stringaction} with metric \eqref{eq:deformed_metric} and $B$ field \eqref{eq:Bfield}, and recast it in the first-order formalism as~\cite{Arutyunov:2009ga,Lloyd:2014bsa} 
\begin{equation}
\label{eq:firstorderaction}
    S=\int\limits_{-\infty}^{+\infty}\de\tau\int\limits_{0}^{R}\de\sigma\left(P_\mu\Xd^\mu+\frac{h^{01}}{h^{00}}C_1+\frac{1}{2\sqrt{-h}h^{00}}C_2\right)\,,
\end{equation}
where $P_{\mu}=\frac{\delta S}{\delta( \partial_\tau X^\mu)}$. Explicitly, the light-cone momenta are given by
\begin{equation}
    P_{\tilde\gamma}=\frac{(\kappa^2-a)\,P_+-P_-}{\kappa^2}\,,\qquad  P_\gamma=\frac{a\,P_++P_-}{\kappa^2}\,,\qquad
    P_+=P_{\tilde\gamma}+P_{\gamma}\,,\qquad P_{-}=\kappa^2 P_{\gamma}-a\,P_+\, , 
\end{equation}
and using the notation $\Xd^\mu=\partial_\tau X^\mu\,,
    \Xp^\mu=T\,\partial_\sigma X^\mu$,
the Virasoro constraints read
\begin{equation}
\label{eq:Virasoro}
\begin{aligned}
    &C_1 \equiv P_\mu \Xp^\mu = 0\,,\\
    &C_2 \equiv G^{\mu\nu}P_\mu P_\nu+G_{\mu\nu}\Xp^\mu \Xp^\nu+2G^{\mu\nu}B_{\nu\rho}P_\mu  \Xp^\rho+G^{\mu\nu}B_{\mu\rho}B_{\nu\sigma}\Xp^\rho \Xp^\sigma =0\, .
\end{aligned}
\end{equation}

Choosing the uniform light-cone gauge-fixing~\cite{Arutyunov:2005hd,Arutyunov:2009ga}
\begin{equation}
\label{eq:lcgauge}
    X^+=\tau\,,\qquad P_{-}=1\, ,
\end{equation}
the solutions to the Virasoro constraints are 
\begin{equation}
    \Xp^-=-P_j\Xp^j\,,\qquad
    P_{+}=-\mathcal{H}(P_j,X^j,\Xp^j)\,,
\end{equation}
where we collectively indicated the transverse coordinate and momenta as $X^j,P_j$, respectively, and $\mathcal{H}(P_j,X^j,\Xp^j)$ is the positive root of a quadratic equation coming from $C_2=0$. Plugging this back into the action~\eqref{eq:firstorderaction} we obtain
\begin{equation}
    S=\int\limits_{-\infty}^{+\infty}\de\tau\int\limits_{0}^{R}\de\sigma\left(P_j\Xd^j-\mathcal{H}(P_j,X^j,\Xp^j)\right)\,,
\end{equation}
which indicates that $\mathcal{H}$ is precisely the worldsheet Hamiltonian in the light-cone gauge.

\vskip 5pt
Introducing the Noether charges corresponding to the $U_{\tilde{\gamma}}(1)$ and $U_\gamma(1)$ isometries 
\begin{equation}
    E=-\int\limits_0^R\de\sigma\,P_{\tilde{\gamma}}\,,\qquad J=\int\limits_0^R\de\sigma\,P_{\gamma}\, ,
\end{equation}
in the gauge~\eqref{eq:lcgauge} we explicitly find 
\begin{equation}
    \mathcal{H}= E-J\,,\qquad
    R=\int\limits_0^R\de\sigma\,P_{-}=\kappa^2 J+a\,\mathcal{H}\,.
\end{equation}
It is worth noting that the volume~$R$ is now fixed, it depends on the gauge parameter~$a$ as it would occur in a $T\overline{T}$ deformation~\cite{Baggio:2018gct,Frolov:2019nrr}. 
In the undeformed case, $\lambda = 0$, there are usually three natural choices for $a$. If $a=0$ we have $\tilde\gamma=X^+=\tau$ and the volume $R$ is totally fixed in terms of the angular momentum $J$.
The other two possibilities are $a=1$, where the gauge fixing implies $\gamma=X^+=\tau$ and consequently $R=E$, and $a=1/2$, which leads to the usual light-cone coordinates $X^+=\frac{1}{2}(\tilde\gamma+\gamma)$ with $R=\frac{1}{2}(J+E)$.
In the presence of the $\lambda$ parameter, similar relations would be obtained by fixing $a=\kappa^2$ and $a=\frac{\kappa^2}{2}$.

With the exception of the level-matching constraint (see e.g.~\cite{Arutyunov:2009ga}) the gauge-fixing \eqref{eq:lcgauge} removes all redundancies from the worldsheet model, and it allows to study it perturbatively as a two-dimensional interacting quantum field theory.

Properly rescaling the coordinates as  
$X^j\to \frac{X^j}{\sqrt{T}}$ and $p_j\to \frac{p_j}{\sqrt{T}}$, the light-cone Hamiltonian can be expanded in powers of $1/\sqrt{T}$ as
\begin{equation}
    \mathcal{H}=\mathcal{H}^{(2)}+ \frac{1}{\sqrt{T}} \mathcal{H}^{(3)} +\frac{1}{T}\mathcal{H}^{(4)}+\dots\,.
\end{equation}

Here, the quadratic Hamiltonian is the same as for the pp-wave background~\cite{Blau:2002dy,Metsaev:2002re}, and reads 
\begin{equation}
    \mathcal{H}^{(2)}=\frac{1}{2}\left[\sum_{j=1}^8(p_j^2+\xp_j^2)-2b(x_1\xp_3-x_3\xp_1-x_2\xp_4+x_4\xp_2)+m^2(x_1^2+x_2^2)+(x_3^2+x_4^2)\right]\,,
    \label{eq:Quadratic_H}
\end{equation}
where we have defined
\begin{equation}
\label{eq:bm}
    b=\frac{1+\lambda^2}{(1+\lambda)^2}\,,\qquad m^2=\frac{1}{\kappa^8}=\left(\frac{1-\lambda}{1+\lambda}\right)^4\,.
\end{equation}
Note that $0\leq m^2\leq1$ for $1\geq \lambda\geq0$. 

For this model there is no cubic Hamiltonian, i.e.
\begin{equation}
   {\cal H}^{(3)}=0\,. 
\end{equation}
Instead, the quartic Hamiltonian takes a fairly cumbersome form, due to the fact that most of the isometries are broken, and the dependence on the deformation parameter is quite non-trivial.
Introducing the coefficients $g_1$, \dots, $g_4$, $b_1$, \dots,  $b_4$, as well as $c_1,\dots,c_4$, that are functions of $\lambda$ and the light-cone gauge parameter $a$,
\begin{equation}
\begin{aligned}
    g_1&=\frac{1}{\kappa^{12}}\frac{\kappa^2(1-34\lambda+\lambda^2)-6a(1-\lambda)^2}{24(1+\lambda)^2} \, ,
        &g_2&=\frac{1}{\kappa^{12}}\frac{\kappa^2(5+22\lambda+5\lambda^2)-6a(1-\lambda)^2}{24(1+\lambda)^2} \, , 
    \\ 
    g_3&=-\frac{a}{4}+\frac{1}{24}\kappa^2 \, ,
    &g_4&=-\frac{a}{4}+\frac{5}{24}\kappa^2 \, ,
\end{aligned}
\end{equation}
\begin{equation}
\begin{aligned}
 b_1&=\frac{\lambda-1}{\kappa^6}\frac{2\kappa^2(1-4\lambda+\lambda^2)-3a(1+\lambda^2)}{6(1+\lambda)^3} \, ,\qquad
        &b_2&=\frac{\lambda -1}{\kappa^6}\frac{\kappa^2(1+8\lambda+\lambda^2)-3a(1+\lambda^2)}{6(1+\lambda)^3} \, ,
        \\
        b_3&=\frac{\lambda-1}{\kappa^6}\frac{2\kappa^2(1-4\lambda+\lambda^2)-a(1+\lambda^2)}{2(1+\lambda)^3} \, ,\qquad
    &b_4&=\frac{1-\lambda}{\kappa^6}\frac{\kappa^2(1-8\lambda+\lambda^2)+a(1+\lambda^2)}{2(1+\lambda)^3} \, ,
\end{aligned}
\end{equation}
\begin{equation}
\begin{aligned}
    c_1&=\kappa^2-3a\,, 
    \qquad
    &c_2&= 2\kappa^2-3a \, ,
    \\
    c_3&=\kappa^2-2a \, ,
    \qquad
    &c_4&=\kappa^2-a \, ,
\end{aligned}   
\end{equation}
we find
\begin{equation}
\label{eq:quarticH}
\begin{aligned}
        &\mathcal{H}^{(4)}=g_1x_1^4+g_2x_2^4+g_3x_3^4+g_4x_4^4+b_1x_1^3x'_3-b_3x_1^2x_3x'_1+b_4x_2^2x_4x'_2-b_2x_2^3x'_4
        \\
    &\quad 
    +\frac{1}{4}\Bigl[ \kappa^2(x_3^2-x_4^2)+\frac{1}{\kappa^6}(x_1^2-x_2^2)\Bigl]\sum_{j=1}^8(p_j^2+\xp_j^2)  +\frac{1}{2\kappa^6}\Bigl[x_1^2(\xp_3^2-p_3^2)-x_2^2(\xp_4^2-p_4^2)\Bigl]
    \\
    &\quad 
    +\frac{a\,b}{2}\Biggl[-x_2x_4^2\xp_4+\frac{1}{\kappa^8}(x_1x_2^2\xp_3-x_2^2x_3\xp_1)+2(p_3^2x_1\xp_3-p_1^2x_3\xp_1)+2(p_2\xp_2+p_4\xp_4)(p_3x_1-p_1x_3)
    \\
    &\quad 
    + 2(p_3x_1-p_1x_3)\sum_{j=5}^8p_j\xp_j+2p_1p_3(x_1\xp_1-x_3\xp_3)+(\xp_1x_3-\xp_3x_1)\Bigl(\sum_{j=1}^8(p_j^2+\xp_j^2)-x_4^2\Bigl) \Biggl] 
    \\
    &\quad
    +\frac{c_4\,b}{2}\Biggl[2(p_4x_2-p_2x_4)\sum_{j=1}^8p_j\xp_j+(x_4\xp_2-x_2\xp_4)\Bigl(\sum_{j=1}^8(p_j^2+\xp_j^2)-\frac{1}{\kappa^8}x_1^2-x_3^2\Bigl)-x_1x_3^2\xp_3\Biggl]
    \\
    &\quad
    +\frac{b}{6}(c_1 x_3^3\xp_1-c_2 x_4^3\xp_2) 
    \\
    &\quad
    +\frac{c_3}{4}\Biggl[\frac{1}{\kappa^{16}}x_1^2x_2^2+x_3^2x_4^2+\frac{1}{\kappa^8}(x_1^2+x_2^2)(x_3^2+x_4^2)-\frac{1}{2}\Bigl(\sum_{j=1}^8(p_j-\xp_j)^2\Bigl)\Bigl(\sum_{j=1}^8(p_j+\xp_j)^2\Bigl)\Biggl] \, .
\end{aligned}
\end{equation}

\section{Dispersion relations and free-particle spectrum}
\label{sec:Hamiltonian}

\vskip 6pt
In this section we determine 
the spectrum of free-particle states and the corresponding dispersion relations, by solving explicitly the free equations of motion. 

The quadratic Hamiltonian describes the free propagation of the transverse modes $(x_1,x_3)$ in AdS, the $(x_2,x_4)$ modes on the sphere and the $(x_5, \dots , x_8)$ coordinates of the torus.
The corresponding free-field equations of motions read
\begin{equation}
\label{eq:free-eom}
\begin{aligned}
 \text{AdS}_3:\qquad & (\square-m^2)x_1+2b\xp_3=0\,, \qquad&& (\square-1)x_3-2b\xp_1=0\,,\\
    \text{S}^3: \qquad & (\square-m^2)x_2-2b\xp_4=0\,, \qquad&&  (\square-1)x_4+2b\xp_2=0\, , \\
    \text{T}^4: \qquad & \square \, x_i =0 \, , \qquad i=5, \dots , 8 \, ,
\end{aligned}
\end{equation}
where $\square=-\partial_\tau^2+\partial_\sigma^2$ is the  two-dimensional flat-space Laplacian and $b$ is defined in \eqref{eq:bm}. 

We solve these equations by taking the decompactification limit, i.e. sending the cylinder radius to infinity (see e.g.~\cite{Arutyunov:2009ga}), and Fourier expanding the coordinates. Precisely, we write
\begin{equation}
    \begin{aligned}
        &x_1=\int \frac{\de p}{\sqrt{2}}\sum_{j=\pm}\Bigl(A_je^{ip\sigma -i\omega_j\tau}{\text a}_j(p)+\bar A_je^{-ip\sigma +i\omega_j\tau}{\text a}_j^\dagger(p)\Bigl)\,, \\
        &x_3=\int \frac{\de p}{\sqrt{2}}\sum_{j=\pm}\Bigl(B_je^{ip\sigma -i\omega_j\tau}{\text a}_j(p)+\bar B_je^{-ip\sigma +i\omega_j\tau}{\text a}_j^\dagger(p)\Bigl)\,,\\
        &x_2=\int \frac{\de p}{\sqrt{2}}\sum_{j=\pm}\Bigl(C_je^{ip\sigma -i\omega_j\tau}{\text b}_j(p)+\bar C_je^{-ip\sigma +i\omega_j\tau}{\text b}_j^\dagger(p)\Bigl)\,, \\
        &x_4=\int \frac{\de p}{\sqrt{2}} \, \sum_{j=\pm}\Bigl(D_je^{ip\sigma -i\omega_j\tau}{\text b}_j(p)+\bar D_je^{-ip\sigma +i\omega_j\tau}{\text b}_j^\dagger(p)\Bigl)\, ,\\
        &x_i = \int \frac{\de p}{\sqrt{2\omega_0}} \Bigl( e^{ip\sigma -i\omega_0\tau} {\text c}_i(p)+ e^{-ip\sigma +i\omega_0\tau}{\text c}_i^\dagger(p)\Bigl)\, \qquad i=5, \dots , 8 \, ,   
    \end{aligned}
    \label{eq:oscillators}
\end{equation}
where, as usual, $\omega_\pm$ and $\omega_0$ are determined in terms of the momentum by requiring these expansions to provide a non-trivial solution of the equations of motion. Specifically, in this case the resulting dispersion relations read\footnote{Similar dispersion relations have been found in the study of other deformed backgrounds~\cite{Georgiou:2022fow, Hoare:2023zti,Borsato:2024sru,Hoare:2025rtl}, in particular in~\cite{Georgiou:2022fow} where the $\lambda$-deformation of $\mathbb{R}\times\text{S}^3$ was studied, but with slightly different conventions.}
\begin{equation}
\label{eq:dispersion}
\omega_\pm(p)=\sqrt{p^2+\frac{1+m^2}{2}\pm\sqrt{4b^2p^2+\frac{(1-m^2)^2}{4}}}\,\,  , \qquad \qquad \omega_0 = |p| \, .
\end{equation}

Plugging expansions \eqref{eq:oscillators} in the equations of motion one finds that the equations for the torus coordinates are automatically satisfied, whereas the rest of equations lead to a linear system that allows to determine the unknown coefficients $A_\pm,B_\pm,C_\pm,D_\pm$, up to an overall normalisation factor. 

Fixing this factor by requiring the ${\text a}_\pm, \text{b}_\pm$ oscillators to be canonically normalised, we eventually find 
\begin{equation}
\label{eq:coefficients}
    A_\pm=-C_\pm=\mp i\frac{f^\mp\sqrt{f^\pm}}{4bp\sqrt{2\omega_\pm\mu}}\,,\qquad
     B_\pm=D_\pm=\frac{\sqrt{f^\pm}}{\sqrt{2\omega_\pm\mu}}\,,
\end{equation}
 where we have introduced
\begin{equation} \begin{gathered}
\label{eq:coefficient_functions}
    f^\pm=\mu\pm M\,,\qquad
    \mu=\sqrt{M^2+16b^2p^2}\,,\qquad
    M=1-m^2=8\lambda \frac{1+\lambda^2}{(1+\lambda)^4} \, .
\end{gathered} \end{equation}

Now, we canonically quantize the system by imposing standard  equal-time commutation relations $[x_j(\sigma),p_k(\sigma')]=i\delta(\sigma-\sigma')\delta_{jk}$
for all the coordinates. With our normalization this implies 
\begin{equation}    
    [{\text a}_{j}(p),{\text a}^\dagger_{k}(q)]=\delta_{jk}\delta(p-q)\,,\qquad  [{\text b}_{j}(p),{\text b}^\dagger_{k}(q)]=\delta_{jk}\delta(p-q)\, , \qquad [{\text c}_j(p), {\text c}_k^\dagger(q) ] = \delta_{jk} \delta(p-q).
    \label{eq:canonical_commutation_rules}
\end{equation}
The oscillators ${\text a}_\pm$, ${\text a}_\pm^\dagger$ and $\text{b}_\pm$, $\text{b}_\pm^\dagger$ are then interpreted as annihilation and creation operators for the transverse coordinates on AdS$_3$ (${\text a}_\pm$, ${\text a}_\pm^\dagger$) and S$^3$ (${\text b}_\pm$, $\text{b}_\pm^\dagger$) respectively, while ($\text{c}_j, \text{c}_j^\dagger$) are the corresponding operators on the torus. Following a notation which is standard in AdS/CFT integrability, we will denote with $Z$ the excitations created by AdS oscillators, with $Y$ those created by sphere oscillators, while $X$ are those on the torus. More precisely, one-particle states are defined as
\begin{equation}
    |Z_{\pm}(p)\rangle=\text{a}^\dagger_{\pm}(p)|0\rangle\,,\qquad
    |Y_{\pm}(p)\rangle=\text{b}^\dagger_{\pm}(p)|0\rangle\,, \qquad 
     |X_j(p)\rangle = \text{c}_j^\dagger |0\rangle \, . 
\end{equation}

\begin{figure}[t]
    \centering    \includegraphics[width=0.45\linewidth]{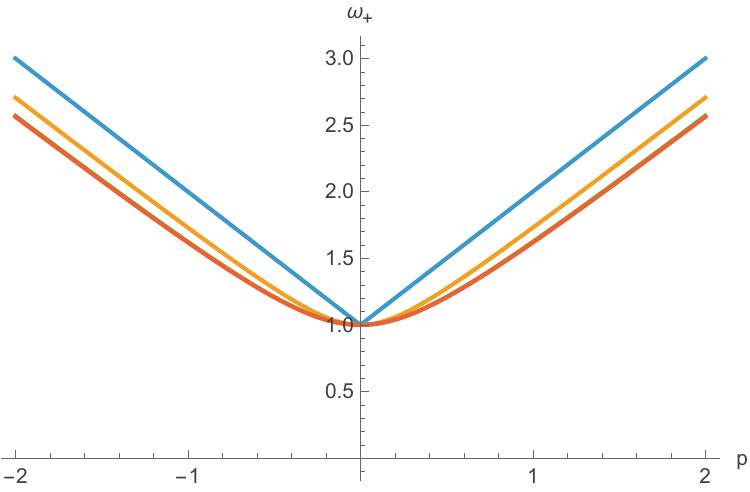}
        \includegraphics[width=0.45
        \linewidth]{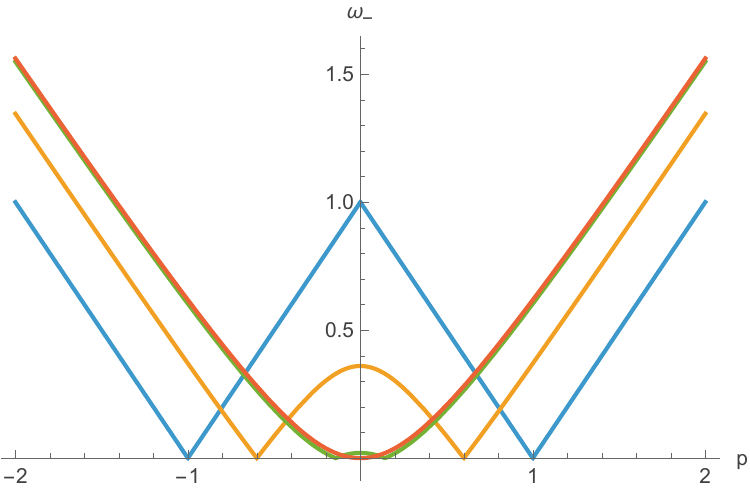}
    \caption{The dispersion relations $\omega_{+}(p)$ (left) and $\omega_{-}(p)$ (right) as functions of the momentum, for fixed values of the deformation parameter~$\lambda$. Different curves correspond to $\lambda=0$ (blue), $\lambda=0.25$ (orange), $\lambda=0.75$ (green), and $\lambda=1$ (red).}
    \label{fig:dispersion}
\end{figure}

\vskip 5pt

In order to determine the particle spectrum, we first need to discuss the analytic properties of the dispersion relations~\eqref{eq:dispersion}. We will focus on the non-trivial dispersions $\omega_\pm(p)$, which are depicted in figure~\ref{fig:dispersion} for different values of the deformation parameter $\lambda$. 

As can be seen in that figure, the $\omega_+(p)$ function has a minimum in $p=0$, such that
\begin{gather}
    \omega_+(p)\big|_{p=0}=1,\qquad 0\leq\lambda\leq1\,.
\end{gather}
It is a smooth function of $p$ for any value of $\lambda$, except  for $\lambda=0$ (the undeformed case) where the dispersion relation is not analytic at $p=0$.

The $\omega_-(p)$ function, instead, exhibits a qualitatively different behaviour as one varies~$\lambda$.
It has in fact two minima at $p=\pm\sqrt{m}$ 
\begin{equation}
\label{eq:min-omegaminus}
    \omega_-(p)\big|_{p=\pm\sqrt{m}}=0,\qquad 0\leq\lambda<1\,,
\end{equation}
where the function is not analytic, and a local maximum at $p=0$ where 
\begin{equation}
\label{eq:max-omegaminus}
    \omega_-(p)\big|_{p=0}=m,\qquad 0\leq\lambda<1\, ,
\end{equation}
and the function is analytic, except for $\lambda =0$. 

In the limit $\lambda\to1$ the two minima~\eqref{eq:min-omegaminus} and the maximum~\eqref{eq:max-omegaminus} merge in a global minimum at $p=0$, where the dispersion is regular.

\vskip 5pt
In order to determine the mass spectrum of the system it is then convenient to distinguish different values of $\lambda$. Moreover, for each range of the deformation parameter it may be necessary to distinguish different branches of $\omega_\pm$ as functions of $p$, which can be identified by values of~$p$ where the group velocity
\begin{equation}
\label{eq:velocity}    v_\pm(p)=\frac{\partial\omega_\pm}{\partial p}=\frac{p}{\omega_\pm}\left(1\pm\frac{4b^2}{\mu}\right)
\end{equation}
is singular. 

\paragraph{Particles of the deformed model $(0<\lambda<1)$.}
As can be seen from figure~\ref{fig:dispersion}, the dispersion $\omega_+(p)$ is perfectly regular, and can be interpret as a non-relativistic particle with mass $m=1$. 

\begin{figure}
    \centering
    \includegraphics[width=0.5\linewidth]{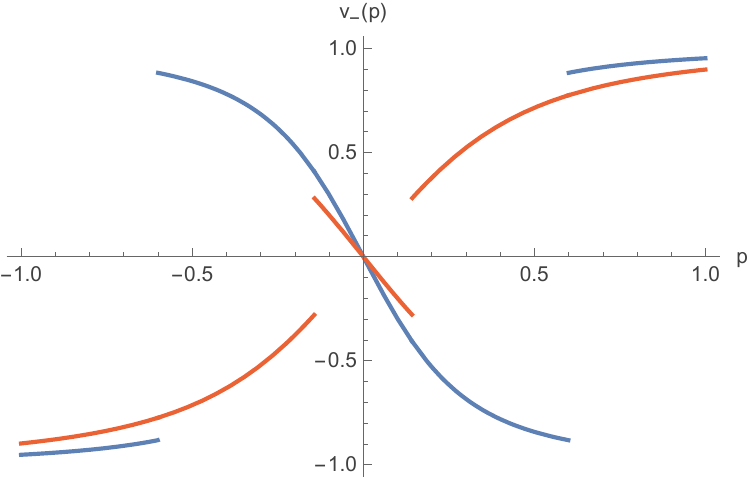}
    \caption{Group velocity $v_-(p)=\partial\omega_{-}/\partial p$ oscillation for $\lambda=0.25$ (blue), and $\lambda=0.75$ (red). The jump discontinuities indicate the presence of different branches in the $\omega_-$ dispersion relation, which can be treated as distinct types of particles.}
    \label{fig:groupvelocity}
\end{figure}

The case of $\omega_-(p)$ is more complicated, as we have to identify three distinct excitations, separated by the points $p=\pm\sqrt{m}$ where the dispersion relation is not analytic.
This can be seen most readily from the plot of the group velocity in figure~\ref{fig:groupvelocity}, which highlights three distinct branches.
None of these excitations has a mass gap. For $p<-\sqrt{m}$ we have a non-relativistic left-moving excitation, and for $p>+\sqrt{m}$ a right-moving one.
Finally, for $|p|<\sqrt{m}$ we have a non-relativistic excitation somewhat reminiscent of a giant magnon~\cite{Hofman:2006xt} or of massless modes in pure-RR $\AdSST$ backgrounds~\cite{Borsato:2014exa}.

\paragraph{Particles of the original model $(\lambda=0)$.}
The spectrum of the undeformed pure-NSNS $\AdSST$ string was studied in the light-cone gauge at tree level in~\cite{Hoare:2013pma} and later at all loops in~\cite{Dei:2018mfl,Borsato:2014exa}. 

\begin{figure}[t]
\centering
    \centering    \includegraphics[width=0.4\linewidth]{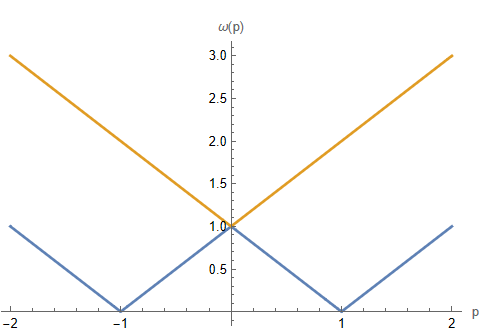}~~~~\includegraphics[width=0.4
        \linewidth]{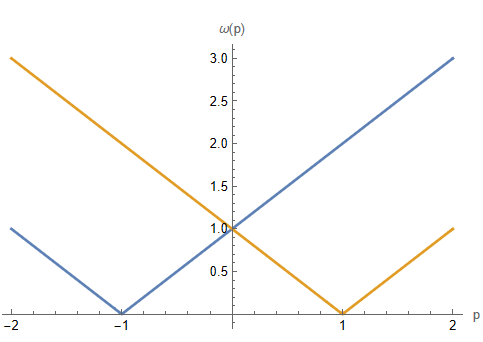}
    \caption{ On the left we plot the dispersion relations \eqref{eq:omega_lambda0} obtained by setting $\lambda =0$ in the general expression \eqref{eq:dispersion}. The yellow line indicates $\omega_+$, whereas the blue line is $\omega_-$. On the right we interpret the particles on the same slope as part of the same branch (see eq.~\eqref{eq:omegaoriginal}), in line with refs.~\cite{Hoare:2013pma,Dei:2018mfl}.}
    \label{fig:NSNS}
\end{figure}

In the present context, setting $\lambda =0$ in \eqref{eq:dispersion} we find
\begin{equation}  \omega_\pm(p)=\sqrt{1+p^2\pm2|p|}=\left|1\pm|p|\right|\, .
     \label{eq:omega_lambda0}
\end{equation}
These functions correspond to the left plots of  figure~\ref{fig:NSNS}. 
Since the different branches of $\omega_\pm$ connect at $p=0$, in order to match the results of the undeformed case it is more natural to re-define them as~\cite{Hoare:2013pma,Dei:2018mfl}  
\begin{equation}
\label{eq:omegaoriginal}
\omega_+(p)=|p+1|\,, \qquad \omega_-(p)=|p-1|\,.
\end{equation}
We are then dealing with two massless particles, whose  dispersion relations have minima  at $p=\pm1$, corresponding to the right plots of figure~\ref{fig:NSNS}. Since  the particles are massless, the dispersion relation is not analytic and the two branches (the left- and right-movers on the worldsheet) should be treated separately.
As explained in~\cite{Dei:2018mfl}, these excitations can be distinguished by their spins in AdS and on the sphere.

\paragraph{Particles of the non-Abelian T-dual model ($\lambda\to1$).}
While at the level of string background the $\lambda\to1$ limit requires some care~\cite{Itsios:2023kma}, in the dispersion relations it is a smooth limit. In fact, we straightforwardly obtain
\begin{equation}
    \omega_\pm(p)=\sqrt{p^2+\frac{1}{2}\pm\sqrt{p^2+\frac{1}{4}}}\,\, .
\end{equation} 
Both expressions are regular for any value of $p$ and satisfy 
\begin{equation}
 \omega_+(p) -  \omega_-(p) = 1\,.
\end{equation}
The $\omega_+(p)$ dispersion exhibits a mass gap and corresponds to a non-relativistic particle excitation with unit mass. Instead, $\omega_-(p)$ is gapless and describes a non-relativistic massless mode.

\section{Tree-level S matrix}
\label{sec:Smatrix}

\vskip 6pt
We are now ready to determine the S matrix governing the two-to-two scattering of the particle modes classified above.  Here, we summarise the fundamental steps for the computation of the tree level S matrix, whereas we present the explicit results in the next section.

Expanding the scattering matrix in the string tension~$T$ 
\begin{equation}
    \mathbf{S}=\mathbf{1}+\frac{i}{T}\mathbf{T}+\mathcal{O}(T^{-2})\, ,
\end{equation}
the first non-trivial contribution is determined by the the $1/T$ interaction Hamiltonian  $\mathcal{H}^{(4)}$ in eq.~\eqref{eq:quarticH}.
Expressing it in terms of the oscillators defined in~\eqref{eq:oscillators}, it takes the following schematic form 
\begin{equation}
\mathcal{H}^{(4)}=
    \int \prod_{n=1}^4 \de p_n \; H^{ijkl}(p_1,p_2,p_3,p_4)\;{\text d}_i^\dagger(p_1){\text d}_j^\dagger(p_2){\text d}_k(p_3){\text d}_l(p_4)\;\delta(p_{\text{in}}-p_{\text{out}})\,\delta(\omega_{\text{in}}-\omega_{\text{out}})\, ,
    \label{eq:quarticH_oscillator}
\end{equation}
where $(\text{d},\text{d}^\dagger)$ generally denote the creation and annihilation operators in  \eqref{eq:canonical_commutation_rules} or combinations thereof.
It then describes two-to-two particle processes with  $p_{\text{in}} = p_1 + p_2$ and $p_{\text{out}} = p_3 + p_4$.

If this were a relativistic theory -- thus with ordinary dispersion relation $\omega = \sqrt{p^2 + m^2}$ -- and we were to consider scattering of  identical mass particles, the energy and momentum conservation would force the two-to-two scattering to be \textit{purely elastic}, that is, we would immediately find $p_3=p_2$ and $p_4=p_1$ (up to re-labeling).

In the present case, however, given the non-trivial dispersion relations \eqref{eq:dispersion}, the kinematical constraints do not necessarily imply the scattering to be elastic.
For instance, let us consider the two-particle scattering process $|\phi_-(p_1)\phi_-(p_2)\rangle\to|\phi_-(p_3)\phi_-(p_4)\rangle$, with $p_1=-p_2 \equiv p $, $|p|<\sqrt{m}$. Imposing the energy and momentum conservation with the energy given by $\omega_-$ in \eqref{eq:dispersion}, we obtain two possible solutions for the outgoing momenta,
\begin{equation}
    p_{1}=-p_2= p\qquad\Rightarrow\qquad
    p_{3}=\begin{cases}
        \pm p\,,&\\
        \pm \sqrt{p^2+4b^2-4b\sqrt{(m-b)^2+p^2}}\,&
    \end{cases},
    \qquad p_4=-p_3.
\end{equation}
While the first solution describes purely elastic processes -- no particle production and conservation of the spectrum of single particle momenta -- the solution in the second line does not.

Similarly, a non-elastic solution appears when we scatter particles of different flavour, that is  $|\phi_+(p_1)\phi_-(p_2)\rangle\to|\phi_+(p_3)\phi_-(p_4)\rangle$ with $|p_2|<\sqrt{m}$, and for any value of $p_1$. In this case, the  transmission process with $p_3=p_1$ and $p_4=p_2$ corresponds to a purely elastic solution, while the ''reflection'' solution $p_3=p_2$ and $p_4=p_1$  corresponds to a non-elastic one. 

\vskip 6pt
This analysis leads to a crucial question: Since the model is classically integrable, the S matrix elements for all the potentially non-elastic processes should identically vanish, once we take into account the precise form of the interactions. 
In the rest of this section we address this point while reporting the explicit results for all two-to-two amplitudes. 

\section{S matrix results}
\label{sec:Smatrixresults}
\vskip 6pt
Scattering amplitudes are given by the $H^{ijkl}$ coefficients in the interaction Hamiltonian~\eqref{eq:quarticH_oscillator}, evaluated at momenta satisfying energy-momentum conservation as implemented by the delta functions. 
Momentum conservation is straightforward and forces $p_4=p_1+p_2-p_3$, whereas the delta function of energy conservation  can be rewritten in terms of its solutions. Solving for instance for $p_3$, it takes the form
\begin{equation}
    \delta(\omega_{\text{in}}-\omega_{\text{out}})=\sum_{j\in\text{solutions}}\frac{\delta(p_3-p_j)}{|\omega'_{\text{out}}(p_j)|} \, , 
\end{equation}
where the set of solutions runs over both the purely elastic and non-elastic ones. 

In order for the particles to scatter, we assume that their group velocities~\eqref{eq:velocity} satisfy
\begin{equation}
    v(p_1)>v(p_2)\,.
\end{equation}
This is particularly important in the case where we have several distinct branches.

For future convenience we introduce the following notation
\begin{gather}
  \omega_0(p_i)=\omega_i^0\,\qquad  \omega^\alpha(p_i)=\omega_i^\alpha=\sqrt{p_i^2+\frac{1+m^2}{2}+\frac{\alpha}{2}\mu_i} \,\, , \, \qquad \alpha=\pm1,
\end{gather}
where the function $\mu_i$ is defined as in \eqref{eq:coefficient_functions}, with the index $i$ referring to the particle's momentum
\begin{gather}
    \mu_i=\sqrt{16\,b^2p_i^2+M^2} \,\, , \ \ \ \text{with} \ \ \ M=1-m^2=8\lambda\frac{1+\lambda^2}{(1+\lambda)^4} \, .
\end{gather}

The S matrix entries assume the following general form
\begin{equation}
    \mathbf{T}|\phi_\alpha(p_1)\psi_\beta(p_2)\rangle=\mathcal{T}^{\phi\psi}_{\alpha\beta}|\phi_\alpha(p_1)\psi_\beta(p_2)\rangle+\mathcal{R}^{\phi\psi}_{\beta\alpha}|\phi_\alpha(p_2)\psi_\beta(p_1)\rangle+\mathcal{I}^{\phi\psi}_{\alpha\beta}|\phi_\alpha(p_1)\psi_\beta(p_2)\rangle 
\end{equation}
for any pair of fields $\phi_{\alpha},\psi_\beta$ with flavours $\alpha,\beta=\pm,j$, where the index $j$ labels the fields on the torus.
Each term corresponds to a different process. We indicate with $\mathcal{T}^{\phi\psi}_{\alpha\beta}$ the transmission process $\phi_\alpha(p_1) \psi_\beta(p_2)\to \phi_{\alpha}(p_1)\psi_\beta(p_2)$, with $\mathcal{R}^{\phi\psi}_{\beta\alpha}$ the reflection $\phi_\alpha(p_1) \psi_\beta(p_2)\to \phi_{\alpha}(p_2)\psi_\beta(p_1)$, and finally with $\mathcal{I}^{\phi\psi}_{\alpha\beta}$ the inelastic amplitudes. 

Once the explicit form of the vertices $H^{ijkl}$ is taken into account\footnote{Given their complicated expressions, we avoid reporting them explicitly. }, we find that the contributions coming from any reflection and inelastic processes are zero, precisely $\mathcal{R}^{\phi\psi}_{\beta\alpha} = \mathcal{I}^{\phi\psi}_{\alpha\beta} =0$ after the delta function integrations.
Note that for particles with energy $\omega_-$, each incoming momentum can belong to any of the three regions shown in figure \ref{fig:groupvelocity}. Consequently, all possible combinations of the three particle types must be considered when explicitly computing the amplitudes.
The vanishing of these contributions has been  verified numerically for various values and combinations of the incoming momenta $p_1$ and $p_2$, as well as for different values of the deformation parameter $\lambda$.
This is a highly non-trivial result, which guarantees that at order $1/T$ the resulting S matrix is purely elastic. In other words, integrability of the string sigma model is preserved at this order.  

The only non-vanishing contributions we are left with are then the transmission amplitudes,
\begin{equation}
\mathbf{T}|\phi_\alpha(p_1)\psi_\beta(p_2)\rangle=\mathcal{T}^{\phi\psi}_{\alpha\beta}|\phi_\alpha(p_1)\psi_\beta(p_2)\rangle \, .
\end{equation}
Below, we list their explicit expressions by distinguishing massive and massless scattering modes.

\subsection{\bf Massive--massive scattering}
\vskip 6pt

Since the explicit form of the scattering amplitudes is fairly involved, it is convenient to organize the results by keeping the couplings in \eqref{eq:quarticH} implicit. 

For the massive-massive scattering of equal particles we find
\begin{equation}
\label{eq:lfunctions}
    \begin{aligned}
        \mathcal{T}^{ZZ}_{\alpha\beta}(p_1,p_2)&=(a\,l^{\alpha\beta}_1+(c_4+c_1)\,l^{\alpha\beta}_2+c_3\,l^{\alpha\beta}_3+(3b_1+b_3)\,l^{\alpha\beta}_4+g_1\,l^{\alpha\beta}_5+g_3\,l^{\alpha\beta}_6+l^{\alpha\beta}_7)J_{\alpha\beta} \, ,
        \\
        \mathcal{T}^{YY}_{\alpha\beta}(p_1,p_2)&=(-c_4\,l^{\alpha\beta}_1+(-a+c_2)l^{\alpha\beta}_2+c_3\,l^{\alpha\beta}_3+(3b_2+b_4)\,l^{\alpha\beta}_4+g_2\,l^{\alpha\beta}_5+g_4\,l^{\alpha\beta}_6-l^{\alpha\beta}_7)J_{\alpha\beta} \, ,
   \end{aligned}
\end{equation}
\vskip 0.6em
\noindent
where the $l_i$ functions depend on the particle flavours and are defined as follows 
\begin{equation}
\begin{aligned}
    l^{\alpha\beta}_1(p_1,p_2)&=8b^6p_1^3p_2^3(\alpha\mu_1+\beta\mu_2)(p_1p_2-\omega_1^\alpha\omega_2^\beta)\, ,
    \\
    l^{\alpha\beta}_2(p_1,p_2)&=b^6p_1^2p_2^2\left[(p_1^2+p_2^2)M+\alpha\mu_1p_2^2+\beta\mu_2p_1^2\right] \, ,
    \\
    l^{\alpha\beta}_3(p_1,p_2)&=
    -\frac{b^4}{8\kappa^{16}}p_1^2p_2^2\Bigl[M^2(1+4\kappa^8)+M^2\kappa^{16}+2\kappa^{16}(16b^2p_1^2+M^2)(16b^2p_2^2+M^2)
    \\ &\quad 
    +\beta\mu_2\kappa^8\left(32b^2p_1^2(1+\kappa^8+2p_2^2\kappa^8)+M^2(3+3\kappa^8+4p_2^2\kappa^8)\right)
    \\ &\quad
    +\alpha\mu_1\kappa^8\left(32b^2p_2^2(1+\kappa^8+2p_1^2\kappa^8)+M^2(3+3\kappa^8+4p_1^2\kappa^8)\right)
    \\ &\quad
    +\alpha \beta\mu_1\mu_2\Bigl(2+\kappa^8\bigl(2+4(p_1^2+p_2^2)\bigr)+\left(2+4p_2^2+4p_1^2(1+4p_2^2)+M^2\right)\kappa^{16}\Bigr)
    \\ &\quad
    +16\alpha\beta p_1p_2\kappa^{16}\omega_1^\alpha\omega_2^\beta\mu_1\mu_2
    \Bigr] \, ,
    \end{aligned}
\end{equation}

\begin{equation}
\begin{aligned}
l^{\alpha\beta}_4(p_1,p_2)&=2b^5p_1^2p_1^2\left[(p_1^2+p_2^2)M-\alpha\mu_1p_2^2-\beta\mu_2p_1^2\right] \, ,
    \\
    l^{\alpha\beta}_5(p_1,p_2)&=3b^4p_1^2p_2^2(M-\alpha\mu_1)(M-\beta\mu_2)\, ,
    \\
    l^{\alpha\beta}_6(p_1,p_2)&=3b^4p_1^2p_2^2(M+\alpha\mu_1)(M+\beta\mu_2) \,,
    \\
    l^{\alpha\beta}_7(p_1,p_2)&=-\frac{b^4}{8\kappa^{14}}p_1^2p_2^2\Bigl[-\beta\mu_2\left(16b^2p_1^2\kappa^{16}+M(\kappa^8-1)(1+\kappa^8+4p_2^2\kappa^8)+M^2\kappa^8(1+\kappa^8)\right)+
    \\ &\quad
    -\alpha\mu_1\left(16b^2p_2^2\kappa^{16}+M(\kappa^8-1)(1+\kappa^8+4p_1^2\kappa^8)+M^2\kappa^8(1+\kappa^8)\right)
    \\ &\quad
    -2\kappa^8\alpha \beta\mu_1\mu_2\left(1-M+\kappa^8+2(p_1^2+p_2^2)(1+\kappa^8)\right)
    \\ &\quad
    -2\left(8b^2(p_1^2+p_2^2)M\kappa^{16}+M^3\kappa^{16}+M^2(1+\kappa^8)\right)
    \\ &\quad
    -64b^2p_1p_2\kappa^8\left((3+\kappa^8)p_1p_2+(\kappa^8-1)\omega_1^\alpha\omega_2^\beta)\right)
    \Bigr] \, .
\end{aligned}
\end{equation}
The global factor in \eqref{eq:lfunctions}
\begin{equation}
      J_{\alpha\beta}(p_1,p_2)=-\frac{1}{2b^4p_1^2p_2^2\left[-\alpha\,p_2\mu_1(-\beta\,\mu_2-4b^2)\omega_1^\alpha+\beta\,p_1\mu_2(-\alpha\,\mu_1-4b^2)\omega_2^\beta\right ]}
\end{equation}
includes also the Jacobian factor coming from the delta function integration
\begin{gather}
    \frac{1}{|\omega'(p_1)-\omega'(p_2)|}=\frac{1}{v(p_1)-v(p_2)} \, , \qquad \text{with} \quad v(p_1)>v(p_2) \, .
\end{gather}

\vskip 5pt
For the mixed processes we find
\begin{equation}
    \begin{aligned}
        \mathcal{T}_{\alpha\beta}^{ZY}(p_1,p_2)&=(a\,s_1^{\alpha\beta}+c_4\,s_2^{\alpha\beta}+c_3\,s_3^{\alpha\beta}+s_4^{\alpha\beta})J_{\alpha\beta} \, ,
        \\
        \mathcal{T}_{\alpha\beta}^{YZ}(p_1,p_2)&=(-c_4\,s_1^{\alpha\beta}-a\,s_2^{\alpha\beta}+c_3\,s_3^{\alpha\beta}-s_4^{\alpha\beta})J_{\alpha\beta} \, ,
    \end{aligned}
\end{equation}
where we have defined the following functions
\begin{equation}
    \begin{aligned}        s^{\alpha\beta}_1(p_1,p_2)&=2\frac{b^6}{\kappa^8}p_1^3p_2^2\left[p_1(M+16b^2p_2^2\kappa^8-M\kappa^8+M^2\kappa^8)+4\beta\,p_2\kappa^8\mu_2(p_1p_2-\omega_1^\alpha\omega_2^\beta)\right] \, , 
        \\
        s^{\alpha\beta}_2(p_1,p_2)&=-2\frac{b^6}{\kappa^8}p_2^3p_1^2\left[p_2(M+16b^2p_1^2\kappa^8-M\kappa^8+M^2\kappa^8)+4\alpha\,p_1\kappa^8\mu_1(p_1p_2-\omega_1^\alpha\omega_2^\beta)\right] \,  ,
        \\
        s^{\alpha\beta}_3(p_1,p_2)&=\frac{b^4}{8\kappa^{16}}p_1^2p_2^2\Bigl[M^2(1-2\kappa^8)+\kappa^{16}\left(M^2-(16b^2p_1^2+M^2)(16b^2p_2^2+M^2)\right)
        \\ &\quad 
        -\beta\mu_2\Bigl(M+(16b^2p_1^2+M^2)\kappa^8+\kappa^{16}\bigl(16b^2p_1^2(1+4p_2^2)+M(-1+M(1+4p_2^2)\bigl)\Bigr)
        \\ &\quad 
        -\alpha\mu_1\Bigl(M+(16b^2p_2^2+M^2)\kappa^8+\kappa^{16}\bigl(16b^2p_2^2(1+4p_1^2)+M(-1+M(1+4p_1^2)\bigl)\Bigr)
        \\ &\quad
        -4\kappa^8\alpha\beta\mu_1\mu_2\Bigl((p_1^2+p_2^2)(1+\kappa^8)+4\kappa^8p_1p_2(p_1p_2-\omega_1^\alpha\omega_2^\beta)\Bigr)
        \Bigr] \, , 
        \\
        s^{\alpha\beta}_4(p_1,p_2)&=\frac{b^4}{8\kappa^{14}}p_1^2p_2^2\Bigl[4\kappa^{8}(p_1^2-p_2^2)\Bigl(4b^2M(1-\kappa^8)-(1+\kappa^8)\alpha\beta\mu_1\mu_2\Bigr)
        \\ &\quad -\beta\mu_2\Bigl(M+\bigl(16b^2p_1^2+M(4p_2^2+M)\bigr)\kappa^8+\bigl(16b^2p_1^2-M(1+4p_2^2-M)\bigr)\Bigr)
        \\ &\quad
        +\alpha\mu_1\Bigl(M+\bigl(16b^2p_2^2+M(4p_1^2+M)\bigr)\kappa^8+\bigl(16b^2p_2^2-M(1+4p_1^2-M)\bigr)\Bigr)
        \Bigr] \, .
    \end{aligned}
\end{equation}

\subsection{\bf Massive-massless scattering}
\vskip 6pt
If one of the two particles involved belongs to the torus directions, due to $SO(4)$ invariance the processes cannot depend on $j$.%
\footnote{While there is no $SO(4)$ invariance on $\text{T}^4$, for the purpose of the perturbative computation of the S~matrix we are treating the torus as if it were~$\mathbb{R}^4$.}

As in the previous case we keep the couplings implicit and we find
\begin{equation}
    \begin{aligned}
        \mathcal{T}_{\alpha}^{ZX}(p_1,p_2)&=(a\,t^\alpha_1+c_3\,t_2^\alpha+t_3^\alpha)J_{\alpha}\, ,
        \qquad
        &\mathcal{T}_{\alpha}^{YX}(p_1,p_2)&=(-c_4\,t^\alpha_1+c_3\,t_2^\alpha-t_3^\alpha)J_{\alpha} \, ,
        \\
        \mathcal{T}_{\ \ \ \alpha}^{XZ}(p_1,p_2)&=\mathcal{T}_{\alpha}^{XZ}(p_2,p_1) \, ,
        \qquad
        &\mathcal{T}_{\ \ \ \alpha}^{XY}(p_1,p_2)&=\mathcal{T}_{\alpha}^{YX}(p_2,p_1) \, ,
    \end{aligned}
\end{equation}
where the following functions have been introduced
\begin{equation}
    \begin{aligned}
         t_1^{\alpha}(p_1,p_2)&=-512b^4p_1^3p_2(p_1p_2-\omega^0_2\omega_1^\alpha) \, ,
         \\
         t^\alpha_2(p_1,p_2)&=32\frac{b^2}{\kappa^8}p_2p_1^2\left[p_2(16b^2p_1^2+M^2)\kappa^8+\alpha\,p_2(1+\kappa^8+4p_1^2\kappa^8)\mu_1-4\alpha\,\kappa^8p_1\mu_1\omega^0_2\omega_1^\alpha\right] \, ,
         \\
         t^\alpha_3(p_1,p_2)&=32\frac{b^2}{\kappa^6}p_1^2p_2^2\left[M(1-\kappa^8)-\alpha(1+\kappa^8)\mu_1\right] \, ,
         \\
         J_{\alpha}(p_1,p_2)&=-\frac{1}{128b^2p_1^2\Bigl[p_1(-4b^2+\alpha\mu_1)\omega^0_2-\alpha\,p_2\mu_1\omega^\alpha_1\Bigr]} \, .
    \end{aligned}
\end{equation}
\subsection{\bf Massless scattering}
\vskip 6pt
Finally, in this case we simply have 
\begin{equation}
    \mathbf{T}|X_{i}(p_1)X_{j}(p_2)\rangle= \mathcal{T}_{ij}^{kl}|X_{k}(p_1)X_{l}(p_2)\rangle\,,
\end{equation}
where $\mathcal{T}_{ij}^{kl}$ is in principle any linear combination of  $SO(4)$ invariant tensors.
However, at tree level it takes the rather simple form
\begin{equation}
    \begin{aligned}
     \mathcal{T}_{ij}^{kl}(p_1,p_2)=\mathcal{T}_{XX}(p_1,p_2)\,\delta_i^k\delta_j^l=
      c_3\,p_1p_2\frac{\omega_1^0\omega_2^0-p_1p_2}{\omega_1^0p_2-\omega_2^0p_1}\,\delta_i^k\delta_j^l \, .
    \end{aligned}
\end{equation}
The means that  only the pure-transmission process is non-zero for massless particles at tree level.
This is the same qualitative behaviour that one sees in the undeformed $\AdSST$ S~matrix~\cite{Hoare:2013pma,Baglioni:2023zsf}.

\vskip 6pt
 It is worth noting that for $\lambda=0$ our results reduce to those of~\cite{Hoare:2013pma,Baglioni:2023zsf} as it must be the case. 

\vskip 5pt
As anticipated above, the appearance of inelastic solutions to the two-body processes could potentially spoil the integrability of the theory. 
The vanishing of the corresponding amplitudes is therefore a non-trivial check that integrability holds at tree-level once the theory is quantized.
The resulting S matrix is manifestly diagonal --- since only transmission processes yield non-zero amplitudes --- and trivially satisfies the classical Yang-Baxter equation
\begin{equation}
    [\mathbf{T}_{12},\mathbf{T}_{13}]+ [\mathbf{T}_{12},\mathbf{T}_{23}] +[\mathbf{T}_{13},\mathbf{T}_{23}]=0 \, ,
\end{equation}
ensuring the tree-level integrability of the theory.

\section{\bf Comments on the S matrix of the non-Abelian T-dual background}
\label{sec:SmatrixNATD}
\vskip 6pt
In the $\lambda\to1$ limit we might expect to find the S-matrix of the non-Abelian T-dual of the $\text{AdS}_3\times\text{S}^3\times\text{T}^4$ background. However, as recalled above and discussed in detail in~\cite{Itsios:2023kma}, this limit is quite delicate and deserves a separate discussion. 

At the level of the metric and the fluxes, it can be performed only in the rescaled set of coordinates~\eqref{eq:NATD_changeofcoordinates}.
In fact, the original $\lambda$-deformed background \eqref{eq:Metric} does not admit a non-Abelian T-dual limit, since there exist no real values of the angles $\tilde\alpha,\tilde\beta,\tilde\gamma$ for which the $SL(2,\mathbb{R})$ manifold is connected to the identity\footnote{At the level of the sigma-model action, the dual model can be obtained by expanding the group element around the identity $g\sim 1+v/k+\mathcal{O}(1/k^2)$ \cite{Sfetsos:2013wia}. Therefore, the condition for the group manifold to be connected to the identity is a necessary condition for the dual model to exist.}. 

The aforementioned limit keeps being problematic and not well-defined also in the S matrix, due to the fact that the quartic Hamiltonian exhibits a singular behavior hidden inside the dependence of some of its coefficients on $\kappa^2= \frac{1+\lambda}{1 - \lambda}$. This is due to the particular choice of  the coefficients in \eqref{eq:ppwave_changeofcoord} and \eqref{eq:ppwave_changeofcoord_LC} required to obtain a quadratic metric in the Brinkmann form. One could then try to consider other geodesics solutions different from \eqref{eq:geodesic}. However, as discussed in section \ref{subsection:near-pp-wave geometry}, the only other allowed solution to the equations is $\beta(\tau)=0$, which is meaningless. 

It is also worth pointing out that one could start from the background fields of the dual model~\cite{Itsios:2023kma}
\begin{equation}
\begin{aligned}
        \de s^2&=\frac{1}{2}\left(\de\rho^2+\frac{\rho^2}{\rho^2-1}(\de\tilde\beta^2-\cosh{\tilde\beta}^2\,\de\tilde\gamma^2)+\de r^2+\frac{r^2}{r^2+1}(\de\beta+\sin{\beta}^2\,\de\gamma^2)\right)+\sum_{i=5}^8\de x_i^2 \, ,
        \\
        B&=\frac{1}{2}\left(\frac{\rho^3}{\rho^2-1}\cosh\tilde\beta\,\de\tilde\beta\wedge\de\tilde\gamma-\frac{r^3}{r^2+1}\sin\beta\,\de\beta\wedge\de\gamma\right)
\end{aligned}
\end{equation}
and find a suitable geodesics for computing the S matrix. However, this turns out to be difficult. In fact, in order to solve the equations one should impose $\tilde{\beta}(\tau)=0$ and $\rho(\tau)=0$, but this solution is forbidden by the constraint $|\rho|>1$, required to preserve the correct metric signature and avoid the singularity. This leads to the conclusion that for $\lambda = 1$ there exists no suitable null geodesics around which we can perform the Penrose limit, at least if we forbid analytic continuations in the metric whose physical meaning is not entirely clear to us. The fact that the quadratic Hamiltonian \eqref{eq:Quadratic_H} is well defined when $\lambda \to 1$, should then be understood as a similarity between the limits and not as the actual T-dual limit.

In conclusion, our present findings about the S-matrix further support the idea that the $\lambda\to 1$ limit alone does not exists, regardless of the fact that the geometry of the dual model can be consistently obtained in this limit.

\section{Conclusions and outlook}
\label{sec:conclusions}

\vskip 6pt
In this work, we have considered a supersymmetric one-parameter $\lambda$-deformation of the  $\AdSST$ string background \cite{Itsios:2023kma}. We have computed the tree-level S matrix for the bosonic excitations, by employing the uniform light-cone gauge to identify the physical degrees of freedom and quantizing the theory via the canonical formalism, after properly defining asymptotic states.

Although the energy dispersion relations kinematically allow also for the existence of inelastic contributions to the scattering processes -- that could possibly spoil the integrability of the theory -- 
we have verified that the non-elastic amplitudes vanish. Therefore, the resulting S matrix is purely elastic and, being also reflectionless, trivially satisfies the classical Yang-Baxter equation, consistently with the integrability of the model.
The absence of non-elastic terms hence constitutes a non-trivial check of our results. The vanishing of potential non-elastic processes, necessary for integrability, was also observed in other deformed models~\cite{Hoare:2023zti,Hoare:2025rtl} as well as in $\text{AdS}_3\times\text{S}^3\times\text{S}^3\times\text{S}^1$ when scattering particles of different masses~\cite{Rughoonauth:2012qd}. Interestingly, there are classically integrable models (in the Lax sense) where integrability-breaking processes appear in the tree-level S~matrix. This is the case of non-Abelian Jordanian deformations~\cite{Borsato:2024sru}.

Furthermore, we have verified that the S matrix is ill-defined in the $\lambda\to 1$ limit, where non-Abelian T-dual geometry is recovered. This supports the idea that such a limit does not exists on its own and can only be applied at the level of the geometry through an appropriate rescaling of the fields and an analytic continuation of the metric.

A natural follow-up of this work would be the introduction of fermion fields to perform the same computation in the mixed-flux regime, as the deformed background also admits non-vanishing Ramond-Ramond (RR) $F_1,F_3,F_5$ fluxes.
Furthermore, it would be of great interest to investigate the model's integrability at the quantum level by directly analyzing the underlying deformed symmetry algebra. A deep understanding of this structure could provide valuable insights towards the bootstrap program, even though for this type of backgrounds such a task remains highly non-trivial.

\vskip 30pt
\noindent
{\bf Acknowledgements}

\vskip 6pt
\noindent
We are grateful to Carlos Nu\~{n}ez for interesting discussions which motivated this work. We also thank Fiona Seibold for useful discussions and comments, as well as Sibylle Driezen and Mario Trigiante for comments on an earlier version of this manuscript.
This work was supported in part by the INFN grant Gauge and String Theory (GAST). AS gratefully acknowledges support from the CARIPARO Foundation under grant No.~68079. 
 
\vskip 40pt

\bibliography{refs}{}

@article{Sfetsos:2013wia,
    author = "Sfetsos, Konstadinos",
    title = "{Integrable interpolations: From exact CFTs to non-Abelian T-duals}",
    eprint = "1312.4560",
    archivePrefix = "arXiv",
    primaryClass = "hep-th",
    reportNumber = "DMUS-MP-13-23, DMUS--MP--13-23",
    doi = "10.1016/j.nuclphysb.2014.01.004",
    journal = "Nucl. Phys. B",
    volume = "880",
    pages = "225--246",
    year = "2014"
}

@article{Itsios:2023kma,
    author = "Itsios, Georgios and Sfetsos, Konstantinos and Siampos, Konstantinos",
    title = "{Supersymmetric backgrounds from {\ensuremath{\lambda}}-deformations}",
    eprint = "2310.17700",
    archivePrefix = "arXiv",
    primaryClass = "hep-th",
    reportNumber = "HU-EP-23/57",
    doi = "10.1007/JHEP01(2024)084",
    journal = "JHEP",
    volume = "01",
    pages = "084",
    year = "2024"
}

@article{Arutyunov:2005hd,
    author = "Arutyunov, Gleb and Frolov, Sergey",
    title = "{Uniform light-cone gauge for strings in AdS(5) x S**5: Solving $SU(1|1)$ sector}",
    eprint = "hep-th/0510208",
    archivePrefix = "arXiv",
    reportNumber = "ITP-UU-05-47, SPIN-05-32, AEI-2005-160",
    doi = "10.1088/1126-6708/2006/01/055",
    journal = "JHEP",
    volume = "01",
    pages = "055",
    year = "2006"
}

@article{Arutyunov:2009ga,
    author = "Arutyunov, Gleb and Frolov, Sergey",
    title = "{Foundations of the AdS$_{5} \times S^{5}$ Superstring. Part I}",
    eprint = "0901.4937",
    archivePrefix = "arXiv",
    primaryClass = "hep-th",
    reportNumber = "ITP-UU-09-05, SPIN-09-05, TCD-MATH-09-06, HMI-09-03",
    doi = "10.1088/1751-8113/42/25/254003",
    journal = "J. Phys. A",
    volume = "42",
    pages = "254003",
    year = "2009"
}

@article{Lloyd:2014bsa,
    author = "Lloyd, Thomas and Ohlsson Sax, Olof and Sfondrini, Alessandro and Stefa{\'n}ski, Jr., Bogdan",
    title = "{The complete worldsheet S matrix of superstrings on AdS$_3 \times$ S$^3 \times$ T$^4$ with mixed three-form flux}",
    eprint = "1410.0866",
    archivePrefix = "arXiv",
    primaryClass = "hep-th",
    reportNumber = "IMPERIAL-TP-OOS-2014-04, HU-MATHEMATIK-2014-21, HU-EP-14-34",
    doi = "10.1016/j.nuclphysb.2014.12.019",
    journal = "Nucl. Phys. B",
    volume = "891",
    pages = "570--612",
    year = "2015"
}

@article{Dei:2018mfl,
    author = "Dei, Andrea and Sfondrini, Alessandro",
    title = "{Integrable spin chain for stringy Wess-Zumino-Witten models}",
    eprint = "1806.00422",
    archivePrefix = "arXiv",
    primaryClass = "hep-th",
    doi = "10.1007/JHEP07(2018)109",
    journal = "JHEP",
    volume = "07",
    pages = "109",
    year = "2018"
}

@article{Hoare:2013pma,
    author = "Hoare, B. and Tseytlin, A. A.",
    title = "{On string theory on $AdS_3 \times S^3 \times T^4$ with mixed 3-form flux: tree-level S-matrix}",
    eprint = "1303.1037",
    archivePrefix = "arXiv",
    primaryClass = "hep-th",
    reportNumber = "IMPERIAL-TP-AT-2013-01, HU-EP-13-10",
    doi = "10.1016/j.nuclphysb.2013.05.005",
    journal = "Nucl. Phys. B",
    volume = "873",
    pages = "682--727",
    year = "2013"
}

@article{Borsato:2014exa,
    author = "Borsato, Riccardo and Ohlsson Sax, Olof and Sfondrini, Alessandro and Stefanski, Bogdan",
    title = "{Towards the All-Loop Worldsheet S Matrix for $AdS_3\times S^3\times T^4$}",
    eprint = "1403.4543",
    archivePrefix = "arXiv",
    primaryClass = "hep-th",
    reportNumber = "IMPERIAL-TP-OOS-2014-01, HU-MATHEMATIK-2014-05, HU-EP-14-12, SPIN-14-11, ITP-UU-14-10",
    doi = "10.1103/PhysRevLett.113.131601",
    journal = "Phys. Rev. Lett.",
    volume = "113",
    number = "13",
    pages = "131601",
    year = "2014"
}

@article{Hofman:2006xt,
    author = "Hofman, Diego M. and Maldacena, Juan Martin",
    title = "{Giant Magnons}",
    eprint = "hep-th/0604135",
    archivePrefix = "arXiv",
    doi = "10.1088/0305-4470/39/41/S17",
    journal = "J. Phys. A",
    volume = "39",
    pages = "13095--13118",
    year = "2006"
}

@article{Hoare:2025rtl,
    author = "Hoare, Ben and Seibold, Fiona K.",
    title = "{Supersymmetry and integrability of the elliptic AdS$_{3}${\texttimes} S$^{3}${\texttimes} T$^{4}$ superstring}",
    eprint = "2507.18548",
    archivePrefix = "arXiv",
    primaryClass = "hep-th",
    doi = "10.1007/JHEP02(2026)139",
    journal = "JHEP",
    volume = "02",
    pages = "139",
    year = "2026"
}

@article{Borsato:2024sru,
    author = "Borsato, Riccardo and Driezen, Sibylle",
    title = "{Particle production in a light-cone gauge fixed Jordanian deformation of AdS5{\texttimes}S5}",
    eprint = "2412.08411",
    archivePrefix = "arXiv",
    primaryClass = "hep-th",
    doi = "10.1103/PhysRevD.111.086010",
    journal = "Phys. Rev. D",
    volume = "111",
    number = "8",
    pages = "086010",
    year = "2025"
}

@article{Hoare:2023zti,
    author = "Hoare, Ben and Retore, Ana L. and Seibold, Fiona K.",
    title = "{Elliptic deformations of the~$\text{AdS}_3 \times \text{S}^3 \times \text{T}^4$ string}",
    eprint = "2312.14031",
    archivePrefix = "arXiv",
    primaryClass = "hep-th",
    reportNumber = "Imperial-TP-FS-2023-02",
    doi = "10.1007/JHEP04(2024)042",
    journal = "JHEP",
    volume = "04",
    pages = "042",
    year = "2024"
}

@article{Georgiou:2022fow,
    author = "Georgiou, George and Sfetsos, Konstantinos",
    title = "{Scattering in integrable pp-wave backgrounds: S-matrix and absence of particle production}",
    eprint = "2208.01072",
    archivePrefix = "arXiv",
    primaryClass = "hep-th",
    reportNumber = "CERN-TH-2022-129",
    doi = "10.1016/j.nuclphysb.2023.116096",
    journal = "Nucl. Phys. B",
    volume = "987",
    pages = "116096",
    year = "2023"
}

@article{Frolov:2019nrr,
    author = "Frolov, Sergey A.",
    title = "{$T\overline T $ Deformation and the Light-Cone Gauge}",
    eprint = "1905.07946",
    archivePrefix = "arXiv",
    primaryClass = "hep-th",
    reportNumber = "TCD-MATH-19-06",
    doi = "10.1134/S0081543820030098",
    journal = "Proc. Steklov Inst. Math.",
    volume = "309",
    number = "1",
    pages = "107--126",
    year = "2020"
}

@article{Baggio:2018gct,
    author = "Baggio, Marco and Sfondrini, Alessandro",
    title = "{Strings on NS-NS Backgrounds as Integrable Deformations}",
    eprint = "1804.01998",
    archivePrefix = "arXiv",
    primaryClass = "hep-th",
    doi = "10.1103/PhysRevD.98.021902",
    journal = "Phys. Rev. D",
    volume = "98",
    number = "2",
    pages = "021902",
    year = "2018"
}

@article{Baglioni:2023zsf,
    author = "Baglioni, Nicola and Polvara, Davide and Pone, Andrea and Sfondrini, Alessandro",
    title = "{On the worldsheet S matrix of the AdS$_{3}$/CFT$_{2}$ mixed-flux mirror model}",
    eprint = "2308.15927",
    archivePrefix = "arXiv",
    primaryClass = "hep-th",
    doi = "10.1007/JHEP05(2024)237",
    journal = "JHEP",
    volume = "05",
    pages = "237",
    year = "2024"
}

@article{Berenstein:2002jq,
    author = "Berenstein, David Eliecer and Maldacena, Juan Martin and Nastase, Horatiu Stefan",
    title = "{Strings in flat space and pp waves from N=4 superYang-Mills}",
    eprint = "hep-th/0202021",
    archivePrefix = "arXiv",
    doi = "10.1088/1126-6708/2002/04/013",
    journal = "JHEP",
    volume = "04",
    pages = "013",
    year = "2002"
}

@article{Blau:2002dy,
    author = "Blau, Matthias and Figueroa-O'Farrill, Jose M. and Hull, Christopher and Papadopoulos, George",
    title = "{Penrose limits and maximal supersymmetry}",
    eprint = "hep-th/0201081",
    archivePrefix = "arXiv",
    reportNumber = "EMPG-02-01, QMUL-PH-02-01",
    doi = "10.1088/0264-9381/19/10/101",
    journal = "Class. Quant. Grav.",
    volume = "19",
    pages = "L87--L95",
    year = "2002"
}

@article{Rughoonauth:2012qd,
    author = "Rughoonauth, Nitin and Sundin, Per and Wulff, Linus",
    title = "{Near BMN dynamics of the AdS(3) x S(3) x S(3) x S(1) superstring}",
    eprint = "1204.4742",
    archivePrefix = "arXiv",
    primaryClass = "hep-th",
    reportNumber = "MIFPA-12-17",
    doi = "10.1007/JHEP07(2012)159",
    journal = "JHEP",
    volume = "07",
    pages = "159",
    year = "2012"
}

@article{Maldacena:1997re,
    author = "Maldacena, Juan Martin",
    title = "{The Large $N$ limit of superconformal field theories and supergravity}",
    eprint = "hep-th/9711200",
    archivePrefix = "arXiv",
    reportNumber = "HUTP-97-A097, HUTP-98-A097",
    doi = "10.4310/ATMP.1998.v2.n2.a1",
    journal = "Adv. Theor. Math. Phys.",
    volume = "2",
    pages = "231--252",
    year = "1998"
}

@article{Maldacena:2000hw,
    author = "Maldacena, Juan Martin and Ooguri, Hirosi",
    title = "{Strings in AdS(3) and SL(2,R) WZW model 1.: The Spectrum}",
    eprint = "hep-th/0001053",
    archivePrefix = "arXiv",
    reportNumber = "CALT-68-2245, CITUSC-99-010, HUTP-99-A027, LBNL-44375, UCB-PTH-99-48, LBL-44375",
    doi = "10.1063/1.1377273",
    journal = "J. Math. Phys.",
    volume = "42",
    pages = "2929--2960",
    year = "2001"
}

@article{Metsaev:2002re,
    author = "Metsaev, R. R. and Tseytlin, Arkady A.",
    title = "{Exactly solvable model of superstring in Ramond-Ramond plane wave background}",
    eprint = "hep-th/0202109",
    archivePrefix = "arXiv",
    reportNumber = "FIAN-TD-02-04",
    doi = "10.1103/PhysRevD.65.126004",
    journal = "Phys. Rev. D",
    volume = "65",
    pages = "126004",
    year = "2002"
}

@article{Beisert:2010jr,
    author = "Beisert, Niklas and others",
    title = "{Review of AdS/CFT Integrability: An Overview}",
    eprint = "1012.3982",
    archivePrefix = "arXiv",
    primaryClass = "hep-th",
    reportNumber = "AEI-2010-175, CERN-PH-TH-2010-306, HU-EP-10-87, HU-MATH-2010-22, KCL-MTH-10-10, UMTG-270, UUITP-41-10",
    doi = "10.1007/s11005-011-0529-2",
    journal = "Lett. Math. Phys.",
    volume = "99",
    pages = "3--32",
    year = "2012"
}

@article{Arutyunov:2007tc,
    author = "Arutyunov, Gleb and Frolov, Sergey",
    title = "{On String S-matrix, Bound States and TBA}",
    eprint = "0710.1568",
    archivePrefix = "arXiv",
    primaryClass = "hep-th",
    reportNumber = "ITP-UU-07-50, SPIN-07-37, TCDMATH-07-15",
    doi = "10.1088/1126-6708/2007/12/024",
    journal = "JHEP",
    volume = "12",
    pages = "024",
    year = "2007"
}

@article{Zamolodchikov:1978xm,
    author = "Zamolodchikov, Alexander B. and Zamolodchikov, Alexei B.",
    editor = "Khalatnikov, I. M. and Mineev, V. P.",
    title = "{Factorized s Matrices in Two-Dimensions as the Exact Solutions of Certain Relativistic Quantum Field Models}",
    reportNumber = "ITEP-35-1978",
    doi = "10.1016/0003-4916(79)90391-9",
    journal = "Annals Phys.",
    volume = "120",
    pages = "253--291",
    year = "1979"
}

@article{Zamolodchikov:1989cf,
    author = "Zamolodchikov, A. B.",
    title = "{Thermodynamic Bethe Ansatz in Relativistic Models. Scaling Three State Potts and Lee-yang Models}",
    reportNumber = "ITEP-89-144",
    doi = "10.1016/0550-3213(90)90333-9",
    journal = "Nucl. Phys. B",
    volume = "342",
    pages = "695--720",
    year = "1990"
}

@article{Hoare:2021dix,
    author = "Hoare, Ben",
    title = "{Integrable deformations of sigma models}",
    eprint = "2109.14284",
    archivePrefix = "arXiv",
    primaryClass = "hep-th",
    doi = "10.1088/1751-8121/ac4a1e",
    journal = "J. Phys. A",
    volume = "55",
    number = "9",
    pages = "093001",
    year = "2022"
}

@inproceedings{Seibold:2024qkh,
    author = "Seibold, Fiona K. and Sfondrini, Alessandro",
    title = "{AdS3 integrability tensionless limits, and deformations: A review}",
    eprint = "2408.08414",
    archivePrefix = "arXiv",
    primaryClass = "hep-th",
    doi = "10.1007/978-3-032-16202-1_5",
    month = "8",
    year = "2024",
    booktitle = "MATRIX Book Series",
    volume = "7"
}

@article{Seibold:2025fnu,
    author = "Seibold, Fiona K. and Sfondrini, Alessandro",
    title = "{Interpolating families of integrable AdS$ _{\mathbf{3}}$ backgrounds}",
    eprint = "2502.07103",
    archivePrefix = "arXiv",
    primaryClass = "hep-th",
    reportNumber = "DESY-25-036",
    doi = "10.1088/1751-8121/addb96",
    journal = "J. Phys. A",
    volume = "58",
    number = "23",
    pages = "235401",
    year = "2025"
}

@article{Hoare:2022asa,
    author = "Hoare, Ben and Seibold, Fiona K. and Tseytlin, Arkady A.",
    title = "{Integrable supersymmetric deformations of AdS$_{3}${\texttimes} S$^{3}${\texttimes} T$^{4}$}",
    eprint = "2206.12347",
    archivePrefix = "arXiv",
    primaryClass = "hep-th",
    reportNumber = "Imperial-TP-AT-2022-02",
    doi = "10.1007/JHEP09(2022)018",
    journal = "JHEP",
    volume = "09",
    pages = "018",
    year = "2022"
}

@article{Seibold:2021lju,
    author = "Seibold, Fiona K. and van Tongeren, Stijn J. and Zimmermann, Yannik",
    title = "{On quantum deformations of AdS$_{3}$ {\texttimes} S$^{3}$ {\texttimes} T$^{4}$ and mirror duality}",
    eprint = "2107.02564",
    archivePrefix = "arXiv",
    primaryClass = "hep-th",
    reportNumber = "Imperial-TP-FS-2021-01, HU-EP-21/19",
    doi = "10.1007/JHEP09(2021)110",
    journal = "JHEP",
    volume = "09",
    pages = "110",
    year = "2021"
}

@article{Sfetsos:2017sep,
    author = "Sfetsos, Konstantinos and Siampos, Konstantinos",
    title = "{Integrable deformations of the $G_{k_1} \times G_{k_2}/G_{k_1+k_2}$ coset CFTs}",
    eprint = "1710.02515",
    archivePrefix = "arXiv",
    primaryClass = "hep-th",
    reportNumber = "CERN-TH-2017-199",
    doi = "10.1016/j.nuclphysb.2017.12.011",
    journal = "Nucl. Phys. B",
    volume = "927",
    pages = "124--139",
    year = "2018"
}

@article{Delduc:2017brb,
    author = "Delduc, Francois and Kameyama, Takashi and Magro, Marc and Vicedo, Benoit",
    title = "{Affine $q$-deformed symmetry and the classical Yang-Baxter  $\sigma$-model}",
    eprint = "1701.03691",
    archivePrefix = "arXiv",
    primaryClass = "hep-th",
    doi = "10.1007/JHEP03(2017)126",
    journal = "JHEP",
    volume = "03",
    pages = "126",
    year = "2017"
}

@article{Sfetsos:2015nya,
    author = "Sfetsos, Konstantinos and Siampos, Konstantinos and Thompson, Daniel C.",
    title = "{Generalised integrable {\ensuremath{\lambda}} - and {\ensuremath{\eta}}-deformations and their relation}",
    eprint = "1506.05784",
    archivePrefix = "arXiv",
    primaryClass = "hep-th",
    doi = "10.1016/j.nuclphysb.2015.08.015",
    journal = "Nucl. Phys. B",
    volume = "899",
    pages = "489--512",
    year = "2015"
}

@article{Hoare:2015gda,
    author = "Hoare, B. and Tseytlin, A. A.",
    title = "{On integrable deformations of superstring sigma models related to $AdS_n \times S^n$ supercosets}",
    eprint = "1504.07213",
    archivePrefix = "arXiv",
    primaryClass = "hep-th",
    reportNumber = "IMPERIAL-TP-AT-2015-02, HU-EP-15-21",
    doi = "10.1016/j.nuclphysb.2015.06.001",
    journal = "Nucl. Phys. B",
    volume = "897",
    pages = "448--478",
    year = "2015"
}

@article{Vicedo:2015pna,
    author = "Vicedo, Benoit",
    title = "{Deformed integrable {\ensuremath{\sigma}}-models, classical R-matrices and classical exchange algebra on Drinfel{\textquoteright}d doubles}",
    eprint = "1504.06303",
    archivePrefix = "arXiv",
    primaryClass = "hep-th",
    doi = "10.1088/1751-8113/48/35/355203",
    journal = "J. Phys. A",
    volume = "48",
    number = "35",
    pages = "355203",
    year = "2015"
}

@article{vanTongeren:2015soa,
    author = "van Tongeren, Stijn J.",
    title = "{On classical Yang-Baxter based deformations of the AdS$_{5}$ {\texttimes} S$^{5}$ superstring}",
    eprint = "1504.05516",
    archivePrefix = "arXiv",
    primaryClass = "hep-th",
    reportNumber = "HU-EP-15-18, HU-MATH-15-05",
    doi = "10.1007/JHEP06(2015)048",
    journal = "JHEP",
    volume = "06",
    pages = "048",
    year = "2015"
}

@article{Hoare:2014oua,
    author = "Hoare, Ben",
    title = "{Towards a two-parameter q-deformation of AdS$_3 \times S^3 \times M^4$ superstrings}",
    eprint = "1411.1266",
    archivePrefix = "arXiv",
    primaryClass = "hep-th",
    reportNumber = "HU-EP-14-44",
    doi = "10.1016/j.nuclphysb.2014.12.012",
    journal = "Nucl. Phys. B",
    volume = "891",
    pages = "259--295",
    year = "2015"
}

@article{Hollowood:2014qma,
    author = "Hollowood, Timothy J. and Miramontes, J. Luis and Schmidtt, David M.",
    title = "{An Integrable Deformation of the $AdS_5 \times S^5$ Superstring}",
    eprint = "1409.1538",
    archivePrefix = "arXiv",
    primaryClass = "hep-th",
    doi = "10.1088/1751-8113/47/49/495402",
    journal = "J. Phys. A",
    volume = "47",
    number = "49",
    pages = "495402",
    year = "2014"
}

@article{Hollowood:2014rla,
    author = "Hollowood, Timothy J. and Miramontes, J. Luis and Schmidtt, David M.",
    title = "{Integrable Deformations of Strings on Symmetric Spaces}",
    eprint = "1407.2840",
    archivePrefix = "arXiv",
    primaryClass = "hep-th",
    doi = "10.1007/JHEP11(2014)009",
    journal = "JHEP",
    volume = "11",
    pages = "009",
    year = "2014"
}

@article{Borsato:2014hja,
    author = "Borsato, Riccardo and Ohlsson Sax, Olof and Sfondrini, Alessandro and Stefanski, Bogdan",
    title = "{The complete AdS$_{3} \times$ S$^3 \times$ T$^4$ worldsheet S matrix}",
    eprint = "1406.0453",
    archivePrefix = "arXiv",
    primaryClass = "hep-th",
    reportNumber = "IMPERIAL-TP-OOS-2014-03, HU-MATHEMATIK-2014-11, HU-EP-14-19, SPIN-14-15, ITP-UU-14-17",
    doi = "10.1007/JHEP10(2014)066",
    journal = "JHEP",
    volume = "10",
    pages = "066",
    year = "2014"
}

@article{Kawaguchi:2014qwa,
    author = "Kawaguchi, Io and Matsumoto, Takuya and Yoshida, Kentaroh",
    title = "{Jordanian deformations of the $AdS_5 x S^5$ superstring}",
    eprint = "1401.4855",
    archivePrefix = "arXiv",
    primaryClass = "hep-th",
    reportNumber = "KUNS-2477, ITP-UU-14-05, SPIN-14-05",
    doi = "10.1007/JHEP04(2014)153",
    journal = "JHEP",
    volume = "04",
    pages = "153",
    year = "2014"
}

@article{Arutyunov:2013ega,
    author = "Arutyunov, Gleb and Borsato, Riccardo and Frolov, Sergey",
    title = "{S-matrix for strings on $\eta$-deformed AdS5 x S5}",
    eprint = "1312.3542",
    archivePrefix = "arXiv",
    primaryClass = "hep-th",
    reportNumber = "ITP-UU-13-31, SPIN-13-23, HU-MATHEMATIK-2013-24, TCD-MATH-13-16",
    doi = "10.1007/JHEP04(2014)002",
    journal = "JHEP",
    volume = "04",
    pages = "002",
    year = "2014"
}

@article{Klimcik:2002zj,
    author = "Klimcik, Ctirad",
    title = "{Yang-Baxter sigma models and dS/AdS T duality}",
    eprint = "hep-th/0210095",
    archivePrefix = "arXiv",
    reportNumber = "IML-02-XY",
    doi = "10.1088/1126-6708/2002/12/051",
    journal = "JHEP",
    volume = "12",
    pages = "051",
    year = "2002"
}

@article{Sfetsos:2014cea,
    author = "Sfetsos, Konstantinos and Thompson, Daniel C.",
    title = "{Spacetimes for $\lambda$-deformations}",
    eprint = "1410.1886",
    archivePrefix = "arXiv",
    primaryClass = "hep-th",
    doi = "10.1007/JHEP12(2014)164",
    journal = "JHEP",
    volume = "12",
    pages = "164",
    year = "2014"
}

@article{Hoare:2014pna,
    author = "Hoare, B. and Roiban, R. and Tseytlin, A. A.",
    title = "{On deformations of $AdS_n$ x $S^n$ supercosets}",
    eprint = "1403.5517",
    archivePrefix = "arXiv",
    primaryClass = "hep-th",
    reportNumber = "IMPERIAL-TP-AT-2014-02, HU-EP-14-10",
    doi = "10.1007/JHEP06(2014)002",
    journal = "JHEP",
    volume = "06",
    pages = "002",
    year = "2014"
}

@article{Delduc:2013qra,
    author = "Delduc, Francois and Magro, Marc and Vicedo, Benoit",
    title = "{An integrable deformation of the $AdS_5 \times S^5$ superstring action}",
    eprint = "1309.5850",
    archivePrefix = "arXiv",
    primaryClass = "hep-th",
    doi = "10.1103/PhysRevLett.112.051601",
    journal = "Phys. Rev. Lett.",
    volume = "112",
    number = "5",
    pages = "051601",
    year = "2014"
}

@article{Seibold:2019dvf,
    author = "Seibold, Fiona K.",
    title = "{Two-parameter integrable deformations of the $AdS_3 \times S^3 \times T^4$ superstring}",
    eprint = "1907.05430",
    archivePrefix = "arXiv",
    primaryClass = "hep-th",
    doi = "10.1007/JHEP10(2019)049",
    journal = "JHEP",
    volume = "10",
    pages = "049",
    year = "2019"
}

@article{Hoare:2022vnw,
    author = "Hoare, Ben and Levine, Nat and Seibold, Fiona K.",
    title = "{Bi-{\ensuremath{\eta}} and bi-{\ensuremath{\lambda}} deformations of {\ensuremath{\mathbb{Z}}}$_{4}$ permutation supercosets}",
    eprint = "2212.08625",
    archivePrefix = "arXiv",
    primaryClass = "hep-th",
    reportNumber = "Imperial-TP-FS-2022-03",
    doi = "10.1007/JHEP04(2023)024",
    journal = "JHEP",
    volume = "04",
    pages = "024",
    year = "2023"
}

@article{Appadu:2017bnv,
    author = "Appadu, Calan and Hollowood, Timothy J. and Price, Dafydd and Thompson, Daniel C.",
    title = "{Yang Baxter and Anisotropic Sigma and Lambda Models, Cyclic RG and Exact S-Matrices}",
    eprint = "1706.05322",
    archivePrefix = "arXiv",
    primaryClass = "hep-th",
    doi = "10.1007/JHEP09(2017)035",
    journal = "JHEP",
    volume = "09",
    pages = "035",
    year = "2017"
}

@article{Demulder:2023bux,
    author = "Demulder, Saskia and Driezen, Sibylle and Knighton, Bob and Oling, Gerben and Retore, Ana L. and Seibold, Fiona K. and Sfondrini, Alessandro and Yan, Ziqi",
    title = "{Exact approaches on the string worldsheet}",
    eprint = "2312.12930",
    archivePrefix = "arXiv",
    primaryClass = "hep-th",
    reportNumber = "NORDITA 2023-083",
    doi = "10.1088/1751-8121/ad72be",
    journal = "J. Phys. A",
    volume = "57",
    number = "42",
    pages = "423001",
    year = "2024"
}

@article{Maurelli:2025iba,
    author = "Maurelli, S. and Noris, R. and Oyarzo, M. and Samtleben, H. and Trigiante, M.",
    title = "{Supersymmetric warped solutions from Type IIB orientifold reduction}",
    eprint = "2504.16822",
    archivePrefix = "arXiv",
    primaryClass = "hep-th",
    doi = "10.1007/JHEP08(2025)013",
    journal = "JHEP",
    volume = "08",
    pages = "013",
    year = "2025"
}

@article{Maurelli:2025ueo,
    author = "Maurelli, S. and Noris, R. and Oyarzo, M. and Trigiante, M.",
    title = "{Warped-AdS$_{3}$ near-horizon geometries from TsT transformations}",
    eprint = "2512.01770",
    archivePrefix = "arXiv",
    primaryClass = "hep-th",
    doi = "10.1007/JHEP03(2026)186",
    journal = "JHEP",
    volume = "03",
    pages = "186",
    year = "2026"
}

@inproceedings{Maurelli:2026dpp,
    author = "Maurelli, Stefano and Noris, Ruggero and Oyarzo, Marcelo and Trigiante, Mario",
    title = "{Deforming ${\rm AdS}_3\times S^3\times T^4$ in Type IIB Supergravity}",
    booktitle = "{25th Hellenic School and Workshops on Elementary Particle Physics and Gravity}",
    eprint = "2604.26854",
    archivePrefix = "arXiv",
    primaryClass = "hep-th",
    month = "4",
    year = "2026"
}
\bibliographystyle{unsrturl}

\end{document}